\documentclass[aps,twocolumn,showpacs,preprintnumbers,nofootinbib,prd,superscriptaddress,groupedaddress,10pt]{revtex4-1}

\makeatletter
\def\l@subsubsection#1#2{}
\def\l@subsubsubsection#1#2{}
\makeatother

\usepackage{graphicx,amssymb,amsmath,amsthm,amsfonts,epsfig,epsf}
\usepackage[usenames]{color}
\usepackage{epstopdf,mathrsfs}

\usepackage{bm}
\usepackage{dcolumn}
\usepackage{latexsym}
\usepackage{rotating}
\usepackage{longtable}

\usepackage{enumerate}
\usepackage{tensor,multirow}
\usepackage{url}
\usepackage[linktocpage]{hyperref}
\usepackage{soul}
\usepackage{cancel}

\def\nn{\nonumber}

\begin{document}

\title{Analytical model for gravitational-wave echoes from spinning remnants}

\author{
Elisa Maggio$^1$,
Adriano Testa$^2$,
Swetha Bhagwat$^1$,
Paolo Pani$^{1,3}$}

\affiliation{${^1}$Dipartimento di Fisica, ``Sapienza'' Universit\`a di Roma \& Sezione INFN Roma1, Piazzale Aldo Moro 
5, 00185, Roma, Italy}
\affiliation{$^{2}$Walter Burke Institute for Theoretical 
Physics,\\\mbox{California Institute of Technology, Pasadena, CA 91125 USA}}
\affiliation{${^3}$Scuola Superiore di Studi Avanzati Sapienza, Viale Regina Elena 291, 00161, Roma, Italy}

\begin{abstract} 
Gravitational-wave echoes in the post-merger signal of a binary coalescence are predicted in various scenarios, 
including near-horizon quantum structures, exotic states of matter in ultracompact stars, and certain deviations from 
general relativity.
The amplitude and frequency of each echo is modulated by the photon-sphere barrier of the 
remnant, which acts as a spin- and frequency-dependent high-pass filter, decreasing the frequency content of each 
subsequent echo. Furthermore, a major fraction of the energy of the echo signal is contained in 
low-frequency resonances corresponding to the quasi-normal modes of the remnant.
Motivated by these features, in this work we provide an analytical gravitational-wave template in the low-frequency 
approximation describing the post-merger ringdown and the echo signal of a spinning ultracompact object. 
Besides the standard ringdown parameters, the template is parametrized in terms of only two physical quantities: 
the reflectivity coefficient and the compactness of the remnant. 
We discuss novel effects related to the spin and to the complex reflectivity, such as a more involved 
modulation of subsequent echoes, the mixing of two polarizations, and the ergoregion instability in case of 
perfectly-reflecting spinning remnants.
Finally, we compute the errors in the estimation of the template parameters with current and future gravitational-wave 
detectors using a Fisher matrix framework. Our analysis suggests that models 
with almost perfect reflectivity can be excluded/detected with current instruments, whereas probing values of 
the reflectivity smaller than $80\%$ at $3\sigma$ confidence level requires future detectors (Einstein Telescope, 
Cosmic Explorer, LISA).
The template developed in this work can be easily implemented to perform a matched-filter based search for echoes and 
to constrain models of exotic compact objects.
\end{abstract}

\maketitle

\section{Introduction}
Gravitational-wave~(GW) echoes in the post-merger signal of a compact binary coalescence might be a smoking gun of 
near-horizon quantum structures~\cite{Cardoso:2016rao,Cardoso:2016oxy,Oshita:2018fqu,Wang:2019rcf}, exotic compact 
objects~(ECOs), exotic states of matter in ultracompact stars~\cite{Ferrari:2000sr,Pani:2018flj,Buoninfante:2019swn}, 
and of modified theories of gravity~\cite{Buoninfante:2019teo,Delhom:2019btt} 
(see~\cite{Cardoso:2017cqb,Cardoso:2017njb,LRR} for 
some recent reviews). Detecting echoes in the GW data of LIGO/Virgo and of future GW 
observatories would allow us to probe the near-horizon structure of compact objects. The absence of echoes in GW 
data could instead place increasingly stronger constraints on alternatives to the black-hole~(BH) paradigm.

Tentative evidence for echoes in the 
combined LIGO/Virgo binary BH events~\cite{Abedi:2016hgu,Conklin:2017lwb} and in the neutron-star binary 
coalescence GW170817~\cite{Abedi:2018npz} have been reported, followed by controversial 
claims about the statistical significance of such 
results~\cite{Abedi:2016hgu,Ashton:2016xff,Abedi:2017isz,Conklin:2017lwb,Westerweck:2017hus,Abedi:2018pst}, and by 
recent negative searches using a more accurate template~\cite{UchikataEchoes} and a morphology-independent 
algorithm~\cite{Tsang2019}.
%
Performing a reliable search for echoes requires developing data analysis techniques as well as constructing 
accurate waveform models. Here we focus on the latter challenge.

While several features of the signal have been understood theoretically~\cite{LRR}, an important open problem is to 
develop templates for echoes that are both accurate and practical for searches in current and future detectors, which 
might complement model-independent~\cite{Abedi:2018npz,Conklin:2017lwb,Conklin:2019fcs} and 
burst~\cite{Tsang:2018uie,Lin:2019qyx,Tsang2019} searches, the latter being independent of the morphology of the echo 
waveform. Furthermore, using an accurate template is crucial for model selection and to discriminate 
the origin of the echoes in case of a detection.
There has been a considerable progress in modeling the echo 
waveform~\cite{Nakano:2017fvh,Mark:2017dnq,Maselli:2017tfq,Bueno:2017hyj,
Wang:2018mlp,Correia:2018apm,Wang:2018gin,UchikataEchoes}, but the approaches adopted so far are not optimal, since 
they are either based on analytical templates not necessarily related to the physical properties of the remnant, or 
rely on model-dependent numerical waveforms which are inconvenient for matched filtered searches and can be 
computationally expensive.
In this paper, we provide an analytical, physically motivated template that is parametrized by the standard ringdown 
parameters plus two physical quantities related to the properties of the exotic remnant. Our template can be 
easily implemented in a matched filter based data analysis.

We extend the recent analytical template of Ref.~\cite{Testa:2018bzd} to include spin 
effects. This is particularly important for various reasons. First, merger 
remnants are typically rapidly spinning (dimensionless spin $\chi\approx 0.7$ in case of nonspinning binaries, 
due to angular-momentum conservation); 
second, the spin might introduce nontrivial effects in the shape and modulation of echoes; finally,
spinning ECOs have a rich phenomenology~\cite{LRR}, for example they might undergo various 
types of instabilities~\cite{1978CMaPh..63..243F,Moschidis:2016zjy,Brito:2015oca,
Cardoso:2007az,Cardoso:2008kj,Chirenti:2008pf,Cardoso:2014sna,Maggio:2017ivp,Maggio:2018ivz}. In particular, 
if an ergoregion instability~\cite{Cardoso:2014sna,Vicente:2018mxl,Brito:2015oca} occurs, the signal would grow 
exponentially in time over a time scale which is generically parametrically longer than the time 
delay between echoes, and it is always much longer than the object's dynamical time scale~\cite{Barausse:2018vdb}.

In this work we use $G=c=1$ units.

\section{Analytical echo template}~\label{sec:Setup}
Reference~\cite{Mark:2017dnq} presents a framework for modeling the echoes from nonspinning ECOs by 
reprocessing the standard BH ringdown (at the horizon) using a transfer function ${\cal K}$, which encodes the 
information about the physical properties of the remnant, such as its reflectivity. Our approach is based on this 
framework, but we extended its scope to gravitational perturbations of \emph{spinning} ECOs. 
Our goal is to model the echo signal \emph{analytically}, following a prescription 
similar to that of the nonspinning case studied in Ref.~\cite{Testa:2018bzd}. The key difference between the present 
work and Ref.~\cite{Testa:2018bzd} is that in the latter the effective potential for 
the perturbations of the Schwarzschild geometry was approximated using a P\"oschl-Teller 
potential~\cite{Poschl:1933zz,Ferrari:1984zz} in order to obtain an analytical solution for BH perturbations. In this 
work, we use a low-frequency approximation to solve Teukolsky's equation analytically. We get an 
analytical transfer function (see Eq.~\eqref{transfer} below) by approximating the BH reflection
(${\cal R}_{\rm BH}$) and transmission (${\cal T}_{\rm BH}$) coefficients. Our final template is provided in a 
ready-to-be-used form in a supplemental {\scshape Mathematica}\textsuperscript{\textregistered} notebook~\cite{webpage}.

\subsection{Background}
%
We consider a spinning compact object with radius 
$r_0$, whose exterior geometry ($r>r_0$) is described by the Kerr metric~\cite{Maggio:2017ivp,Abedi:2016hgu,
Wang:2018gin, Barausse:2018vdb}. 
Unlike the case of spherically symmetric spacetimes, the absence of Birkhoff's theorem in axisymmetry 
does not ensure that the vacuum region outside a spinning object is described by the Kerr geometry. This implies that 
the multipolar structure of a spinning ECO might be different from that of a Kerr 
BH~\cite{Raposo:2018xkf,Barcelo:2019aif}. Nevertheless, for perturbative solutions to the vacuum Einstein's equation 
that admit a smooth BH limit, all multipole moments of the external spacetime approach those
of a Kerr BH in the high-compactness regime~\cite{Raposo:2018xkf} (for specific examples, 
see~\cite{Pani:2015tga,Uchikata:2015yma,
Uchikata:2016qku,Yagi:2015hda,Yagi:2015upa,Posada-Aguirre:2016qpz}).

Therefore, in Boyer-Lindquist coordinates, the line element at $r>r_0$ reads
\begin{eqnarray}
ds^2&&=-\left(1-\frac{2Mr}{\Sigma}\right)dt^2+\frac{\Sigma}{\Delta}dr^2-\frac{
4Mr}{\Sigma}a\sin^2\theta d\phi dt  \nn \\
&+&{\Sigma}d\theta^2+
\left[(r^2+a^2)\sin^2\theta +\frac{2Mr}{\Sigma}a^2\sin^4\theta
\right]d\phi^2\,,\label{Kerr}
\end{eqnarray}
In the above equation $\Sigma=r^2+a^2\cos^2\theta$ and $\Delta=r^2+a^2-2M r = (r - r_{+})(r - r_{-})$, where $r_{\pm}=M 
\pm \sqrt{M^2-a^2}$; $M$ and $J \equiv aM \equiv \chi M^2$ are the total mass and angular momentum of the object 
respectively.

The properties of the object's interior and surface can be parametrized in terms of boundary conditions 
at $r=r_0$, in particular by a complex and (possibly) frequency-and-spin-dependent
reflection coefficient, ${\cal R}$~\cite{Mark:2017dnq,Maggio:2017ivp}. Motivated by models of microscopic corrections at 
the horizon scale, in the following we focus on the case
\begin{equation}
 r_0 = r_+(1+\epsilon) \qquad 0<\epsilon\ll 1\,, \label{epsilon-def}
\end{equation}
where $r_+$ is the location of the would-be horizon. 
We fix $r_0$ (or, equivalently, $\epsilon$), by requiring the location of the 
surface to be at a proper length $\delta\ll M$ from $r_+$, where
\begin{eqnarray}
\delta = \int_{r_{+}}^{r_0} dr \,\sqrt{g_{rr}} |_{\theta=0} \,.
\end{eqnarray}
This implies
\begin{eqnarray}
\epsilon\simeq\sqrt{1-\chi^{2}}\frac{\delta^{2}}{4r_+^2}\,,
\label{proper}
\end{eqnarray}
in the $\delta/M\ll1$ limit.

We shall use $M$, $\chi$, and $\delta/M$ to parametrize the background geometry, and ${\cal R}$ to model 
the boundary conditions for perturbations.

\subsection{Linear perturbations}
%
Scalar, electromagnetic and gravitational perturbations in the exterior Kerr
geometry are described by Teukolsky's master equations~\cite{Teukolsky:1972my,Teukolsky:1973ha,Teukolsky:1974yv}, 
the radial solution of which shall be denoted by 
$_{s}R_{lm}(r,\omega)$ (see Appendix~\ref{app:lowfreq}). 

It is convenient to make a 
change of variables by introducing the Detweiler's 
function~\cite{1977RSPSA.352..381D,Maggio:2018ivz}
\begin{equation}
 \tilde\Psi = \Delta^{s/2} \sqrt{r^2+a^2} \left[\alpha \
_{s}R_{lm}+\beta \Delta^{s+1} \frac{d_{s}R_{lm}}{dr}\right]\,,\label{DetweilerX}
\end{equation}
where $\alpha$ and $\beta$ are certain radial 
functions~\cite{1977RSPSA.352..381D,Maggio:2018ivz} that satisfy the following relation
\begin{equation}
 \alpha^2 - \alpha' \beta \Delta^{s+1} + \alpha \beta' \Delta^{s+1} - \beta^2 \Delta^{2s+1} V_S = \text{constant} \,.
\end{equation}
The radial potential $V_S$ is defined below in Eq.~\eqref{eq:vs}, and $s=0,\pm1,\pm2$ for scalar, electromagnetic and 
gravitational perturbations, respectively.
By introducing the tortoise coordinate $x$, defined as
\begin{equation}
 \frac{dx}{dr}=\frac{r^2+a^2}{\Delta}\,,\label{tortoise}
\end{equation}
Teukolsky's master equation becomes
\begin{equation}
 \frac{d^2 \tilde\Psi}{dx^2}- V(r,\omega) \tilde\Psi=\tilde S \,. \label{master}
\end{equation}
Here $\tilde S$ is a source term and the effective potential reads as
\begin{equation}
 V(r,\omega)=\frac{U\Delta }{(r^2+a^2)^2}+G^2+\frac{dG}{dx}\,, \label{potentialmaster}
\end{equation}
with
\begin{eqnarray}
G &=& \frac{s(r-M)}{r^2+a^2}+\frac{r \Delta}{(r^2+a^2)^2} \,, \\
U &=& V_S+\frac{2\alpha' + (\beta' \Delta^{s+1})'}{\beta \Delta^s} \,, \\
V_S &=& -\frac{1}{\Delta}\left[K^2-is\Delta'K+\Delta(2isK'-\lambda_s)\right] \,,\label{eq:vs}
\end{eqnarray}
and $K=(r^2+a^2)\omega-am$. 
The prime denotes a derivative with respect to $r$. Remarkably, the functions $\alpha$ and
$\beta$ can be chosen such that the resulting potential~\eqref{potentialmaster} is purely
real~\cite{1977RSPSA.352..381D,Maggio:2018ivz}. Although the choice of $\alpha$ and $\beta$ is not unique, 
$\tilde{\Psi}$ evaluated at the asymptotic infinities ($x\to\pm\infty$) remains unchanged up to a phase. Therefore, 
the energy and angular momentum fluxes are not affected~\cite{Chandra}.

The asymptotic behavior of the potential is 
\begin{equation}
 V\to\left\{\begin{array}{ll}
             -\omega^2  & \quad{\rm as}\,\, x\to+\infty\\
             -k^2	& \quad{\rm as}\,\, x\to-\infty
            \end{array}
\right.\,,
\end{equation}
where $k=\omega-m\Omega$ and $\Omega=a/(2Mr_+)$ is the 
angular velocity at the event horizon of a Kerr BH.
%

\subsection{Transfer function}
%
Equation~\eqref{master} is formally equivalent to the static scalar case~\cite{Mark:2017dnq} and can be solved 
using Green's function techniques. At asymptotic infinity, we require the solution of 
Eq.~\eqref{master} to be an outgoing wave, $\tilde 
\Psi(\omega,x\to\infty)\sim \tilde Z^{+}(\omega) e^{i\omega x}$. Similarly to what shown in Ref.~\cite{Mark:2017dnq} 
we have
\begin{equation}
 \tilde Z^{+}(\omega) = \tilde Z_{\rm BH}^{+}(\omega)+ {\cal K}(\omega) \tilde 
Z_{\rm BH}^{-}(\omega)\,. \label{signalomega}
\end{equation}
In the above equation, $\tilde Z_{\rm BH}^{\pm}$ are the responses of a 
Kerr BH (at infinity and near the horizon, for the plus and minus 
signs, respectively) to the source $\tilde S$, i.e. 
\begin{equation}
 \tilde Z_{\rm BH}^{\pm}(\omega) = \frac{1}{W_{\rm BH}}\int_{-\infty}^{+\infty}dx {\tilde S 
\tilde \Psi_{\mp} }\,, \label{ZBH}
\end{equation}
where $\tilde \Psi_\pm$ are two independent solutions of the homogeneous equation associated 
to Eq.~\eqref{master} such that
\begin{equation}\label{Psip}
\tilde \Psi_+(\omega, x) \sim \begin{cases}
 \displaystyle 
e^{+i \omega x} & \text{ as } x \to + \infty\\ 
 \displaystyle  
 B_{\rm out}(\omega)e^{+i k x}  +  B_{\rm in}(\omega) e^{- i k x} & \text{ as } x \to 
- \infty
\end{cases} \,,\\
\end{equation}
\begin{equation}
\tilde \Psi_-(\omega, x) \sim \begin{cases}
 \displaystyle 
 A_{\rm out}(\omega)e^{+i \omega x}  +  A_{\rm in}(\omega) e^{-i \omega x} & \text{ as } x \to + \infty \\ 
 \displaystyle  
 e^{-i k x} & \text{ as } x \to - \infty \\
\end{cases} \,,
\end{equation}
and $W_{\rm BH}=\frac{d\tilde \Psi_+}{dx}\tilde \Psi_- -\tilde 
\Psi_+ \frac{d\tilde \Psi_-}{dx}=2 i k B_{\rm out}$ is the Wronskian of the solutions $\tilde{\Psi}_{\pm}$.
The details of the ECO model are all contained in the transfer function, which is formally the same 
as in Ref.~\cite{Mark:2017dnq}, namely\footnote{A heuristic derivation 
of Eq.~\eqref{transfer} guided by an analogy with the geometrical optics is 
provided in Refs.~\cite{Testa:2018bzd,LRR} for the static case.}\footnote{{The phase 
$e^{-2 i k x_0}$ in Eq.~\eqref{transfer} accounts for waves that travel from the potential barrier to $x=x_0$ 
and return to the potential barrier after being reflected at the surface. Notice that the definition of the transfer 
function and, in turn, various subsequent formulas could be simplified by defining $\bar{\cal R}\equiv{\cal R} 
e^{-2 i k x_0}$. We choose to keep the notation of Ref.~\cite{Mark:2017dnq} instead.}}
\begin{equation}
{\cal K}(\omega)=\frac{{\cal T}_{\rm BH} {\cal R}(\omega){e^{-2 i k x_0}}}{1-{\cal 
R}_{\rm BH} {\cal R}(\omega){e^{-2 i k x_0}}}\,, \label{transfer}
\end{equation}
where ${\cal T}_{\rm BH} = 1/B_{\rm out}$ and ${\cal R}_{\rm BH} = B_{\rm 
in}/B_{\rm out}$ are the transmission and reflection coefficients for waves 
coming from the \emph{left} of 
the photon-sphere potential barrier~\cite{Vilenkin:1978uc,Chandra,NovikovFrolov}. 
The Wronskian relations imply that $|{\cal R}_{\rm BH}|^2+\frac{\omega}{k}|{\cal T}_{\rm 
BH}|^2=1$ for any frequency and spin~\cite{PhysRevD.71.124016}. 

Finally, the reflection coefficient at the surface of the object, ${\cal R}(\omega)$, is
defined such that
\begin{equation}\label{eq:boundary}
 \tilde \Psi \sim e^{-i k(x-x_0)} + {\cal R}(\omega) e^{ik(x-x_0)} \qquad \text{as} \ x \sim x_{0}\, ,
\end{equation}
where $|x_0|\gg M$.

\subsection{The BH reflection coefficient in the low-frequency approximation}
In Appendix~\ref{app:lowfreq} we solve Teukolsky's equation analytically in the low-frequency limit for 
gravitational 
perturbations. We obtain an analytical expression for ${\cal R}_{\rm BH}$ 
which is accurate when $\omega M\ll1$ (we call this the low-frequency approximation hereon). 
This is the most interesting regime for echoes, since they are obtained by reprocessing the 
post-merger ringdown signal~\cite{Mark:2017dnq}, whose frequency content is initially 
dominated by the BH fundamental QNM ($\omega\lesssim\omega_{\rm 
QNM}\sim 0.5/M$) and subsequently decreases in time. The photon-sphere barrier acts as a 
high-pass filter and consequently the frequency content 
decreases for each subsequent echo. Hence, a low-frequency approximation becomes increasingly more accurate at late 
times. We quantify this in Sec.~\ref{spectrogram}.

From the analysis in Appendix~\ref{app:lowfreq}, we find that
\begin{equation}
 {\cal R}_{\rm BH}^{\rm LF}= \sqrt{1+Z} e^{i\Phi}\,, \label{eq:lowfreqAppx}
\end{equation}
where ``LF'' stands for ``low frequency'', and 
\begin{equation}
 Z = 4 Q \beta_{sl}\prod_{n=1}^{l} \left(1 + \frac{4 Q^2}{n^2}\right)
\left[\omega (r_+ - r_-)\right]^{2l+1}  \label{amplfactor}
\end{equation}
coincides with Starobinski's result for the reflectivity of a Kerr BH~\cite{Starobinskij2} (for the sake of 
generality we wrote it for spin-$s$ perturbations), $\sqrt{\beta_{sl}}=\frac{(l-s)! (l+s)!}{(2l)! (2l+1)!!}$, and $Q = 
-k\frac{r_+^2 + a^2}{r_+ - r_-} $. 
The matched asymptotic expansion presented in Appendix~\ref{app:lowfreq} 
allows us to extract also the phase $\Phi=\Phi(\omega,\chi)$. Note that $\Phi$ depends on the choice of an
arbitrary constant in the definition of the tortoise coordinate (see Eq.~\eqref{tortoise}). However, as one would 
expect, this freedom in the choice of $x$ does not affect $\mathcal{K}(\omega)$, since it cancels out in the product 
$\mathcal{R} \mathcal{R}_{{\rm BH}}$.

Furthermore, the phase of $\mathcal{R}(\omega)$ and $\mathcal{R}_{{\rm BH}}$ depends also on the choice of the 
radial perturbation function, but the combination $\mathcal{R} 
\mathcal{R}_{{\rm BH}}$ which enters the transfer function~\eqref{transfer} does not depend on this choice, as expected; 
see Sec.~\ref{sec:phase} for more details.

At low frequencies $\mathcal{R}_{{\rm BH}}$ takes the form described in 
Eq.~\eqref{eq:lowfreqAppx}, while in the high-frequency regime ${\cal R}_{\rm BH}\sim e^{-2\pi\omega/\kappa_H}$, where 
$\kappa_H=\frac{1}{2} (r_+ -r_-)/(r_+^2+a^2)$ is the surface gravity of a Kerr BH~\cite{Neitzke:2003mz,Harmark:2007jy}. 
We, then, use a Fermi-Dirac interpolating function to smoothly connect the two regimes:
\begin{equation}
 {\cal R}_{\rm BH}(\omega,\chi)={\cal R}_{\rm BH}^{\rm 
LF}(\omega,\chi)\frac{\exp{\left(\frac{-2\pi\omega_{R}}{\kappa_H}\right)}+1}{\exp{\left(\frac{2\pi(|\omega|-\omega_{R})}
{\kappa_H}\right)}+1}\,, \label{RBH}
\end{equation}
where $\omega_{R}$ is the real part of the fundamental QNM of a 
Kerr BH with spin $\chi$. For $|\omega| \ll \omega_R$ the reflection coefficient reduces to ${\cal R}_{\rm 
BH}^{\rm LF}$, whereas it is exponentially suppressed when $|\omega|\gg\omega_R$.

The transition between low and high frequencies is phenomenological and not unique, but the choice of the interpolating 
function is not crucial since high-frequency ($\omega\gtrsim\omega_R$) signals are not trapped within the 
photon-sphere and hence are not reprocessed.

\subsection{Modeling the BH response at infinity}
We model the BH response at infinity using the fundamental $l=m=2$ QNM; extensions to multipole modes are 
straightforward. We consider a generic linear combination of two independent polarizations, 
namely~\cite{Berti:2005ys,Buonanno:2006ui}
\begin{eqnarray}
 Z_{\rm BH}^{+} (t)&\sim& \theta(t - t_0) \left(\mathcal{A}_+ \cos(\omega_R 
t+\phi_+)\right.\nonumber\\
&&\left.+i\mathcal{A}_\times \sin(\omega_R t+\phi_\times)\right) e^{-t/\tau}\,, \label{ZBHplus}
\end{eqnarray}
so that $\Re[Z_{\rm BH}^{+}]$ and $\Im[Z_{\rm BH}^{+}]$ are the two ringdown polarizations, $h_+(t)$ and $h_\times(t)$, 
respectively. In the above relation, $\tau=-1/\omega_I$ is the damping time, 
${\cal A}_{+,\times}\in\Re$ and $\phi_{+,\times}\in\Re$ are respectively the amplitudes and the phases of the two 
polarizations, and $t_0$ parametrizes the starting time of the ringdown. 
Note that Eq.~\eqref{ZBHplus} is the most generic expression for the fundamental $l=m=2$ ringdown 
and requires that ${\cal A}_{+,\times}$ and $\phi_{+,\times}$ are four independent parameters. The most relevant case 
of a binary BH ringdown is that of circularly polarized waves~\cite{Buonanno:2006ui}, 
which can be obtained from Eq.~\eqref{ZBHplus} by setting ${\cal A}_{+}={\cal A}_\times$ and $\phi_+=\phi_\times$. 
In the following we provide a template for the generic expression~\eqref{ZBHplus}, but for simplicity in the analysis 
we shall restrict to ${\cal A}_\times=0$, i.e. to linearly polarized waves.

Given that the BH response is in the time domain, the frequency-domain waveform can be obtained through a Fourier 
transform,
\begin{equation}
\tilde Z_{\rm BH}^{\pm}(\omega) = \int_{- \infty}^{+ \infty} \frac{dt}{\sqrt{2 
\pi}} Z_{\rm BH}^{\pm}(t) e^{i \omega t},
\end{equation}
which at infinity simplifies to 
\begin{eqnarray}\label{eq:bhtemplateINF}
 \tilde Z_{\rm BH}^{+}(\omega) \sim \frac{e^{i \omega t_0}}{2\sqrt{2\pi}}  &&\left( 
\frac{\alpha_{1+} {\cal A}_+ -\alpha_{1\times} {\cal A}_\times}{\omega - \omega_{\rm QNM}} \right.\nn\\
&&\left.+  \frac{\alpha_{2+} {\cal 
A}_+ +\alpha_{2\times} {\cal A}_\times}{\omega + \omega_{\rm QNM}^*} 
\right)\,,
\end{eqnarray}
where $\omega_{\rm QNM}=\omega_R+i\omega_I$, $\alpha_{1+,\times}=ie^{-i(\phi_{+,\times} + t_0\omega_{\rm QNM})}$, and 
$\alpha_{2+,\times}=-\alpha_{1+,\times}^*$.

\subsection{Modeling the BH response at the horizon}
Moving to the near-horizon BH response, we focus on $Z_{\rm BH}^{-}$, which is the quantity reprocessed by the transfer 
function (see Eq.~\eqref{signalomega}). Here we generalize the approach of Ref.~\cite{Testa:2018bzd}, which considered 
a source localized near the surface of the ECO. 
Inspection of Eq.~\eqref{ZBH} reveals that $Z_{\rm BH}^{-}(\omega)$ in general contains the same poles in the complex 
frequency plane as $Z_{\rm BH}^{+}(\omega)$. Therefore, the near-horizon response at intermediate times can be written 
as in Eq.~\eqref{eq:bhtemplateINF} with different amplitudes and phases.
Nonetheless, for a given source, $Z_{\rm BH}^{+}(\omega)$ and $Z_{\rm BH}^{-}(\omega)$ are related to each other in a 
non-trivial fashion through Eq.~\eqref{ZBH}.
Let us assume that the source has support only in the interior of the object, i.e.,
on the left of the effective potential barrier, where $V\approx -k^2$. 
This is a reasonable assumption, since the source in the exterior can hardly perturb the spacetime within the cavity 
and therefore its contribution is expected to be subdominant (for example see Refs.~\cite{Wang:2019rcf,Oshita:2019sat}).
In this case, it is easy to show that
\begin{equation}
 \tilde Z_{\rm BH}^-=\frac{{\cal R}_{\rm BH}}{{\cal T}_{\rm BH}}\tilde Z_{\rm BH}^+ +\frac{1}{{\cal T}_{\rm BH}W_{\rm 
BH}} 
\int_{-\infty}^{+\infty} dx\,{\tilde S}e^{ikx}\,.
\end{equation}
Using Eqs.~\eqref{ZBH} and \eqref{Psip} and the fact that ${\tilde S}$ has support only where $V\approx -k^2$, the 
above equation can be written as
\begin{equation}
 \tilde Z_{\rm BH}^-=\frac{{\cal R}_{\rm BH}\tilde Z_{\rm BH}^+ + \tilde {\cal Z}_{\rm BH}^+}{{\cal T}_{\rm BH}}\,, 
\label{ZBHminusTOT}
\end{equation}
where $\tilde {\cal Z}_{\rm BH}^+$ is the BH response at infinity to an \emph{effective} source ${\tilde {\cal 
S}}(\omega,x)={\tilde S}(\omega,x) e^{2ikx}$ within the cavity. As such, the ringdown part of $\tilde {\cal Z}_{\rm 
BH}^+$ can also be generically written as in Eq.~\eqref{eq:bhtemplateINF} but with different 
amplitudes, phases, and starting time. Note that Eq.~\eqref{ZBHminusTOT} is valid for any source (with support only in 
the cavity) and for any spin.

Two interesting features of Eq.~\eqref{ZBHminusTOT} are noteworthy. First, in the final response 
(Eq.~\eqref{signalomega}) the term ${\cal T}_{\rm BH}$ in the denominator of Eq.~\eqref{ZBHminusTOT} cancels out with 
that in the transfer function, Eq.~\eqref{transfer}.
Second,
{Eq.~\eqref{ZBHminusTOT} does not require an explicit modeling of the source. More precisely, although both 
$\tilde {Z}_{\rm BH}^+$ and $\tilde {\cal Z}_{\rm BH}^+$ are linear in the source, they can be written as in 
Eq.~\eqref{eq:bhtemplateINF} which depends on amplitudes, phases, and starting time of the ringdown. Thus, 
Eq.~\eqref{ZBHminusTOT} can be computed \emph{analytically} using 
the expressions for ${\cal R}_{\rm BH}$ and ${\cal T}_{\rm BH}$.}

\begin{table*}[th]
 \begin{tabular}{ll}
 \hline
 \hline
  $\delta$  					& proper distance of the surface from the horizon 
radius $r_+$  \\
  {${\cal R}(\omega)$} 
  & reflection coefficient at the surface (located at 
$x=x_0(\delta)$ in 
tortoise coordinates) \\ 
  \hline
  $M$					& total mass of the object \\
  $\chi$				& angular momentum of the object \\
  ${\cal A}_{+,\times}$		        & amplitudes of the two polarizations of the BH ringdown 
at infinity  \\
  $\phi_{+,\times}$			& phases of the two polarizations of the BH ringdown at 
infinity  \\
  $t_0$					& starting time of the BH ringdown at infinity \\
  \hline
 \end{tabular}
 \caption{Parameters of the ringdown$+$echo template presented in this work. 
 The parameter $\delta$ and the (complex) function
 {$\cal R(\omega)$} characterize 
 the ECO. The remaining $7$ parameters characterize the most generic fundamental-mode BH ringdown. For circularly 
polarized waves (${\cal A}_{+}={\cal A}_{\times}$ and $\phi_{+}=\phi_{\times}$) or for linearly polarized waves 
(for example $A_\times=0$), the number of ordinary BH ringdown parameters reduces to $5$.
 } \label{tab:template}
\end{table*}

\subsection{Ringdown$+$echo template for spinning ECOs} \label{sec:template}
We can now put together all the ingredients previously derived.
The ringdown$+$echo template in the frequency domain is given by Eq.~\eqref{signalomega}.
As already mentioned, by substituting Eq.~\eqref{ZBHminusTOT} in the transfer 
function ${\cal K}$ [Eq.~\eqref{transfer}], the dependence on ${\cal T}_{\rm BH}$ of the second term in 
Eq.~\eqref{signalomega} disappears and one needs to model only the reflection coefficient ${\cal R}_{\rm BH}$. Clearly, 
for ${\cal R}=0$ one recovers a single-mode BH ringdown template in the frequency domain. 

The extra term in Eq.~\eqref{signalomega} associated with the echoes reads
\begin{eqnarray}
  {\cal K}\tilde Z_{\rm BH}^- &=&\frac{{\cal R} {e^{-2 i k x_0}}}{1-{\cal R}_{\rm BH}{\cal R}  {e^{-2 i k x_0}}} 
\left({\cal 
R}_{\rm BH}
\tilde Z_{\rm BH}^+ +\tilde {\cal Z}_{\rm BH}^+\right) \,, \label{template0}
\end{eqnarray}
where ${\cal R}_{\rm BH}$ is given by Eq.~\eqref{RBH} and $\tilde Z_{\rm BH}^+$ is given by 
Eq.~\eqref{eq:bhtemplateINF}. {Note that, while ${\cal R}_{\rm BH}$
depends} on the arbitrary constant associated to the tortoise coordinate [Eq.~\eqref{tortoise}], the 
final expression Eq.~\eqref{template0} does not, as expected.

Remarkably, Eq.~\eqref{template0} does not depend \emph{explicitly} on the source, the latter being entirely 
parametrized in terms of $Z_{\rm BH}^+$ and ${\cal Z}_{\rm BH}^+$, i.e. in terms of the amplitudes of BH ringdown.
Since the two terms in Eq.~\eqref{template0} are additive, in the following we shall focus only on the first one, in 
which the source is parametrized in terms of $Z_{\rm BH}^+$ only. Namely, we shall use
\begin{eqnarray}
  {\cal K}\tilde Z_{\rm BH}^- &=&\frac{{\cal 
R}_{\rm BH}{\cal R} {e^{-2 i k x_0}}}{1-{\cal R}_{\rm BH}{\cal R} {e^{-2 i k x_0}}}
\tilde Z_{\rm BH}^+  \,. \label{template0bis}
\end{eqnarray}
A discussion on the expressions for 
$\tilde {\cal 
Z}_{\rm BH}^+$ in terms of different sources is given in Appendix~\ref{app:response}.
Thus, the final template depends on 
$7$ ``BH'' parameters ($M$, $\chi$, ${\cal A}_{+,\times}$, $\phi_{+,\times}$, $t_0$) plus two ``ECO'' 
quantities: $\delta$ (which sets the location of the surface or, equivalently, the compactness of the object) and the 
complex, frequency-dependent reflection coefficient ${\cal R}(\omega)$, 
see Table~\ref{tab:template}. 

The template presented above is publicly available in a ready-to-be-used supplemental {\scshape 
Mathematica}\textsuperscript{\textregistered} notebook~\cite{webpage}.

\begin{figure*}[th]
\centering
\includegraphics[width=0.48\textwidth]{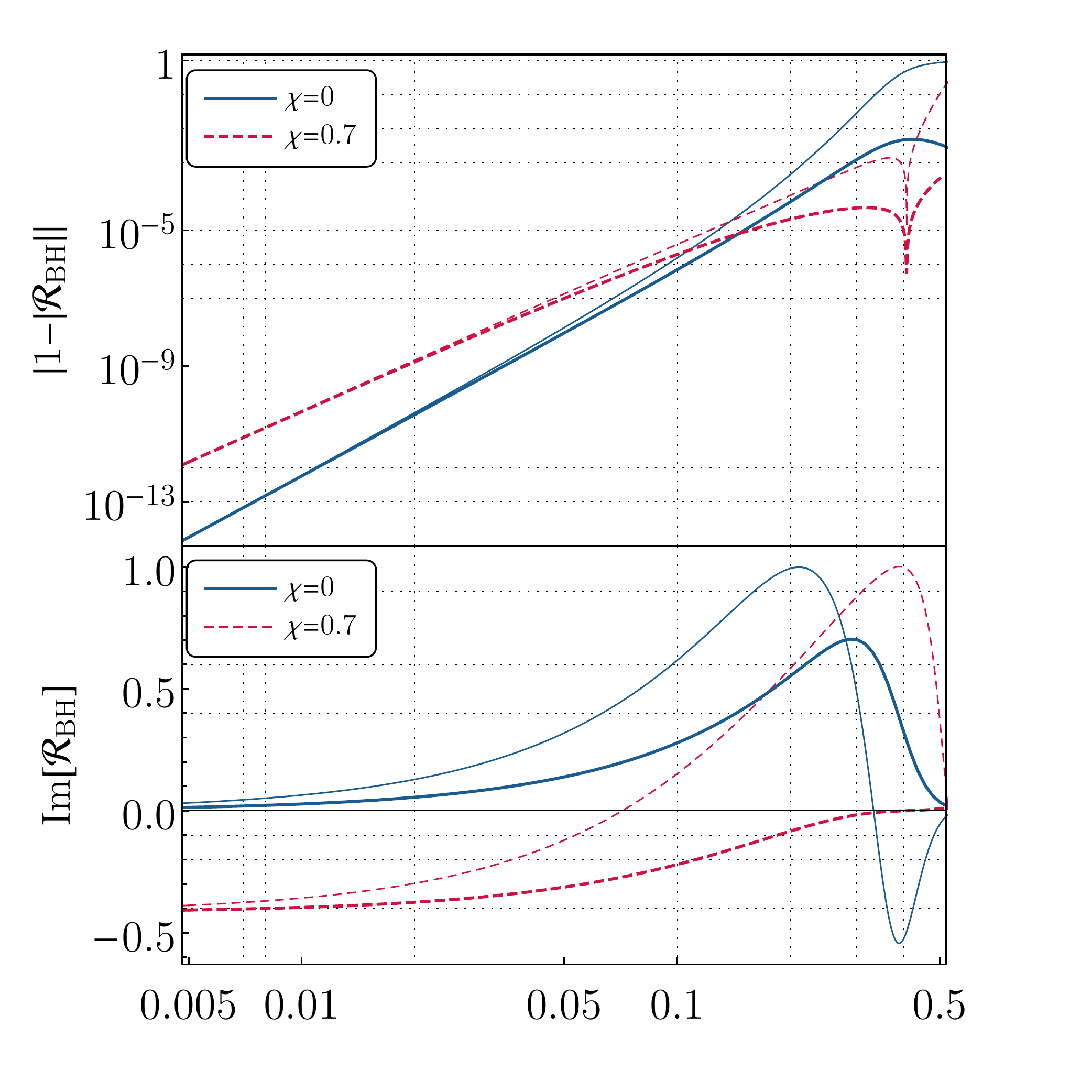}
\includegraphics[width=0.48\textwidth]{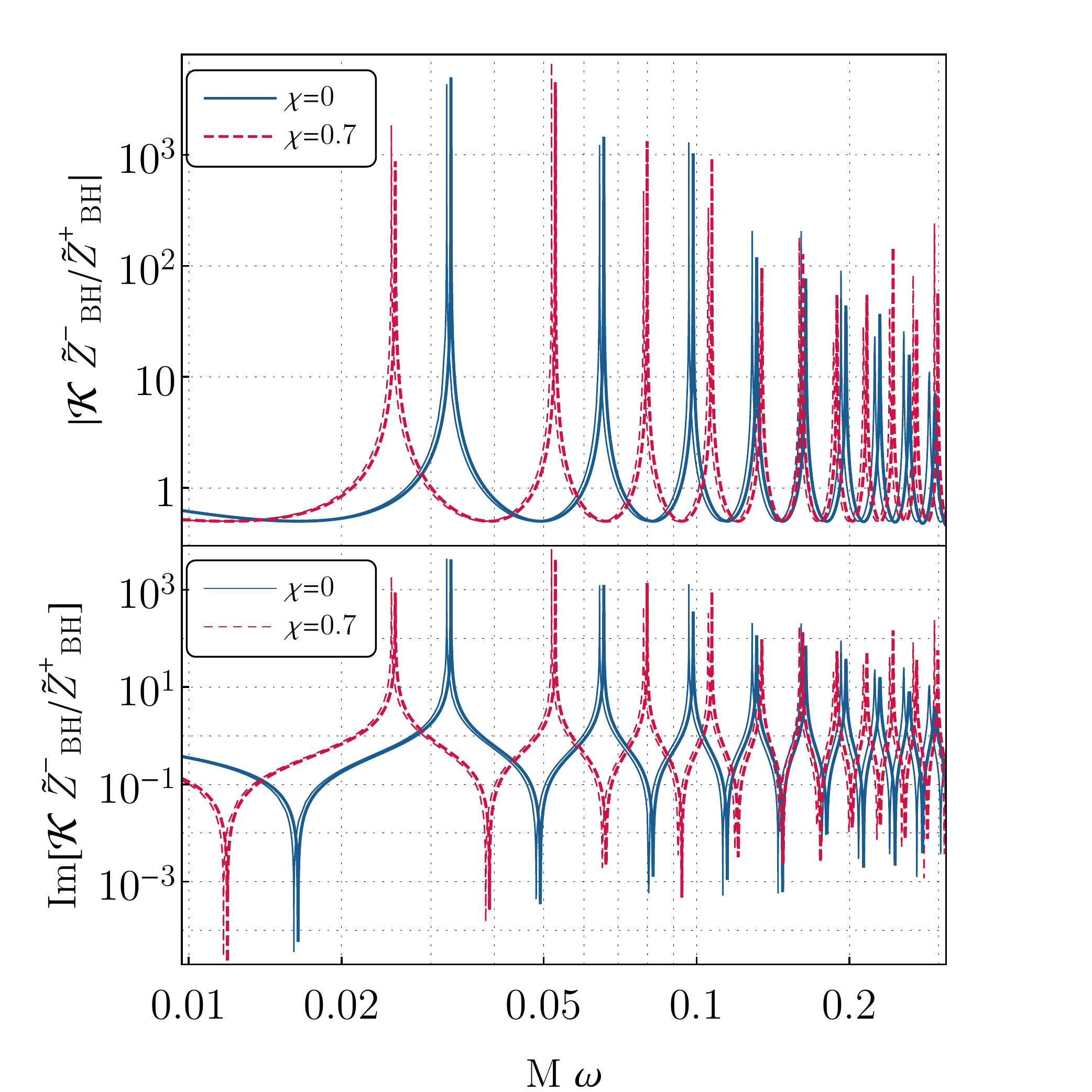}
\caption{Comparison between our analytical template (thick curves) and the result of a 
numerical integration of Teukolsky's equation (thin curves) for $\chi=0$ and {$\chi=0.7$}. {Left} panels: the 
(complex) BH reflection coefficient. Note that the dip in the spinning case corresponds to the threshold of 
superradiance, i.e. $|{\cal R}_{\rm BH}|^2>1$ when $\omega<m\Omega$. {Right} panels: the absolute value ({top}) 
and 
the imaginary part ({bottom}) of the ECO response ${\cal K}\tilde Z_{\rm BH}^-/\tilde Z_{\rm BH}^+$ as functions of 
the frequency. For all panels we chose $l=m=2$ and, for the {right} panels, {$\delta/M=10^{-10}$} and 
${\cal 
R}=1$.} 
\label{fig:comparison_signal_FD}
\end{figure*}

\section{Properties of the template}
\subsubsection{Comparison with the numerical results}
Our analytical template agrees very well with the exact numerical results at low frequency. A representative example is 
shown in Fig.~\ref{fig:comparison_signal_FD}, where we compare
the (complex) BH reflection coefficient ${\cal R}_{\rm BH}$ ({left} panels)
and the echo template ({right} panels) against the result of a numerical integration of Teukolsky's equation. In the 
{right} panels of Fig.~\ref{fig:comparison_signal_FD} we show the quantity ${\cal K}Z_{\rm BH}^-$, normalized by the 
standard BH response $Z_{\rm BH}^+$; since $Z_{\rm BH}^-$ is proportional to $Z_{\rm BH}^+$, the final result is 
independent of the specific BH response. The agreement (both absolute value and imaginary part) is very good at low 
frequencies, whereas deviations are present in the transition region where $\omega M\sim 0.1$. Crucially, the 
low-frequency resonances --~which dominate the response~\cite{Conklin:2017lwb,Conklin:2019fcs}~-- are properly 
reproduced.

{Notice that the agreement between analytics and numerics improves as $\delta\to0$, since the ECO QNMs are at 
lower frequency (for moderate spin) in this regime and our framework is valid.}
For technical reasons we were able to produce numerical results up to $\delta=10^{-10}M$, 
but we expect that the agreement would improve significantly for more realistic (and significantly smaller) values, 
when $\delta$ is of the order of the Planck length.

{To quantify the agreement, we compute the overlap 
\begin{equation}
 {\cal O}=  \frac{|\langle \tilde h_A | \tilde h_N\rangle|}{\sqrt{|\langle \tilde h_N | \tilde h_N\rangle| |\langle 
\tilde 
h_A | \tilde h_A\rangle}|} \label{overlap}
\end{equation}
between the analytical signal, $\tilde h_A$, and the numerical one, $\tilde h_N$, where the inner product is defined as
\begin{equation}
 \langle \tilde X | \tilde Y\rangle \equiv 4 \Re  \int_{0}^{\infty} \frac{\tilde X(f) \tilde Y^*(f)}{S_n(f)} df\,,
\end{equation}
(or in a certain frequency range), $S_n$ is the detector's noise spectral density, and $f=\omega/(2\pi)$ is the GW 
frequency.}

{When $|{\cal R}|\sim 1$ the presence of very 
high and narrow resonances makes a quantitative comparison challenging, since a 
slight displacement of the resonances (due for instance to finite-$\omega$ truncation errors)
deteriorates the overlap. 
For instance, for a representative case shown in Fig.~\ref{fig:comparison_signal_FD} ($\delta=10^{-10}M$, $\chi=0.7$, 
and ${\cal R}=1$) the overlap is excellent (${\cal O}\gtrsim0.999$) when the integration is performed before the first 
resonance, but it quickly reduces to zero after that. 
To overcome this issue, we compute the overlap in the case in which the resonances are less pronounced, as it is 
the case when $|{\cal R}|<1$. Let us consider 
$M=30\,M_\odot$, $\chi=0.7$, $\delta=10^{-10}M$, and the aLIGO noise spectral 
density~\cite{zerodet}.
For ${\cal R}=0.9$, the overlap in the range $f\in(20,100)\,{\rm Hz}$ (whose upper end roughly corresponds to the 
limit $\omega M\sim0.1$ beyond which the low-frequency approximation is not accurate) is ${\cal O}=0.48$. 
This 
small value is mostly due to a small displacement of the resonances. Indeed, by shifting the mass of the analytical 
waveform by only $1.6\%$, the overlap increases significantly, ${\cal O}=0.995$.
For ${\cal R}=0.8$ and in the same conditions, we get ${\cal O}\approx0.8$ 
without mass shift and ${\cal O}\gtrsim0.999$ with the same mass shift as above with the mass 
shift indicated above. As $\delta \rightarrow 0$, the shift in the mass decreases since the exact resonant frequencies
are better reproduced.} 


\subsubsection{Time-domain echo signal: modulation and mixing of the polarizations}
The time-domain signal can be computed through an inverse Fourier transform,
\begin{equation}
h(t) = \frac{1}{\sqrt{2 \pi}}\int_{- \infty}^{+ \infty} d\omega 
\tilde{Z}^{+}(\omega) e^{-i \omega t}\,, \label{inverseFT}
\end{equation}
where $\Re[h(t)]$ and $\Im[h(t)]$ are the two polarizations of the wave, respectively.

\begin{figure*}[th]
 \centering
  \includegraphics[width=1\textwidth]{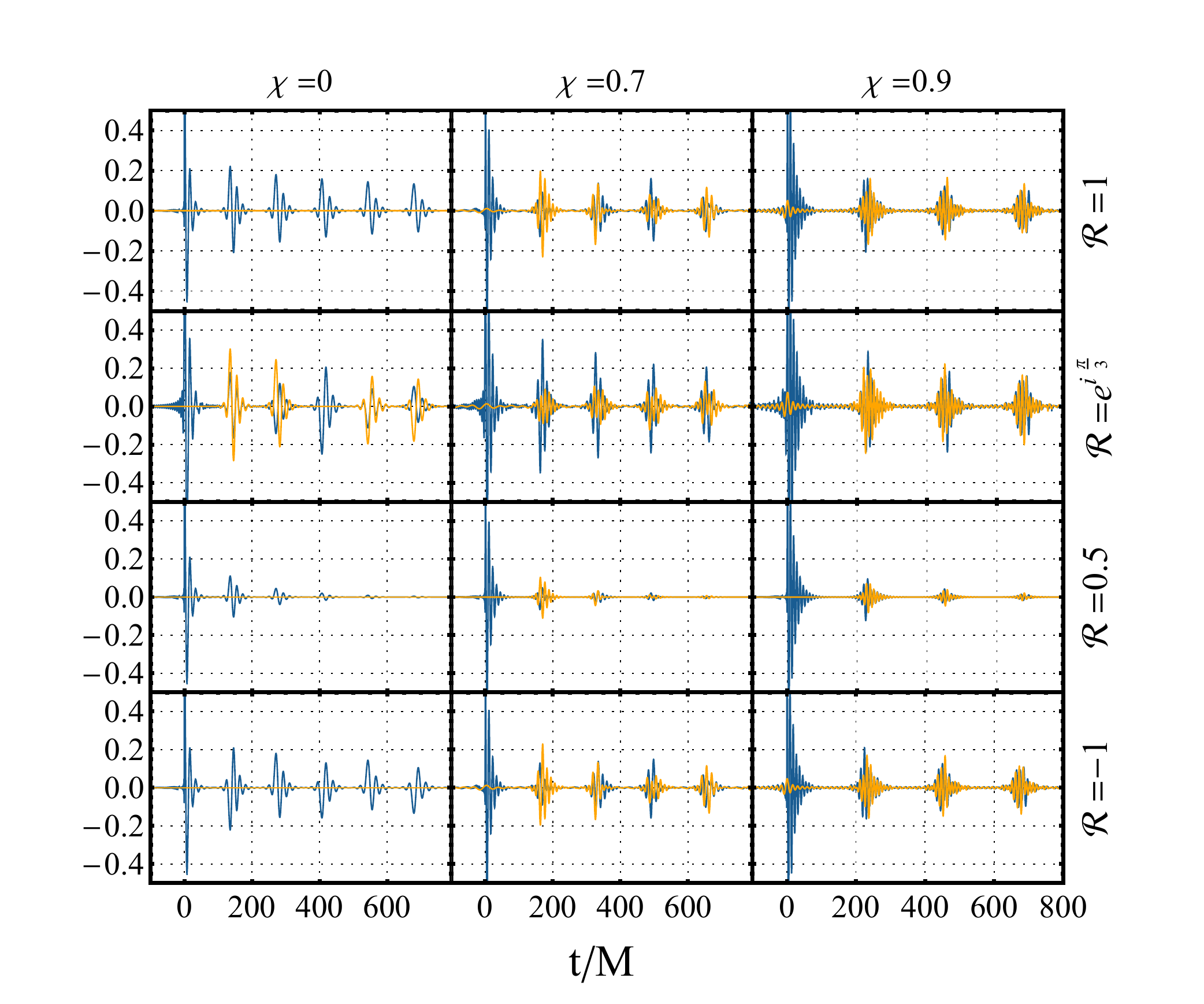}
 \caption{
 Examples of the gravitational ringdown$+$echo template in the time domain for 
 different values of ${\cal R}(\omega)={\rm const}$, and object's spin $\chi$. We consider $\delta/M=10^{-7}$.
 We plot the real (blue curve) and the imaginary (orange curve) parts of the waveform, corresponding to the plus and 
 cross polarization, respectively (note that the ringdown signal is purely plus-polarized, i.e. ${\cal A}_\times=0$). 
 Each waveform is normalized to the peak of $|\Re [h(t)]|$ during the 
 ringdown (the peak is not shown in the range of the $y$ axis to better visualize the subsequent echoes).
 Additional waveforms are provided online~\cite{webpage}.
 }
 \label{fig:template_time}
 \end{figure*}

In Fig.~\ref{fig:template_time} we present a 
representative slideshow of our template for different values of {$\cal 
R$} and spins. For simplicity, we consider $\delta/M=10^{-7}$ and
${\cal R(\omega)}={\rm const}$ (but generically 
complex). The time-domain waveform contains all the features previously reported 
for the echo signal, in particular amplitude and frequency 
modulation
\cite{Cardoso:2016rao,Cardoso:2016oxy,Cardoso:2017cqb,Cardoso:2017njb,Testa:2018bzd}.

In addition, the spin of the object and the phase of the reflectivity coefficient introduce novel effects, such 
as a nontrivial amplitude modulation of subsequent echoes. This is 
mostly due to the spin-and-frequency dependence of the phase of ${\cal R}_{\rm BH}$ and 
${\cal R}$. The effect of the spin can be seen by comparing the left 
column ($\chi=0$) of Fig.~\ref{fig:template_time} with the middle ($\chi=0.7$) and the right columns ($\chi=0.9$). 
Note that the phase of each subsequent echo depends on the combination ${\cal R}{\cal R}_{\rm 
BH}$, i.e., on the combined action of the reflection by the surface and by the BH barrier. Thus, 
phase inversion~\cite{Abedi:2016hgu,Wang:2018gin,Testa:2018bzd} of each echo relative to the 
previous one occurs whenever ${\cal R}{\cal R}_{\rm BH}\approx -1$ for low frequencies (cf. Sec.~\ref{sec:phase} for 
more details).

Furthermore, note that the first, the second, and the fourth row of Fig.~\ref{fig:template_time} all correspond to 
perfect reflectivity, $|{\cal R}|=1$, but their echo structure is different: in other words, a phase term in 
${\cal R}$ introduces a nontrivial echo pattern. To the best of our knowledge this effect was neglected in the 
previous analyses.

%

%
%

As shown in Fig.~\ref{fig:template_time} the time-domain signal can contain both plus and cross polarizations, even if 
the initial ringdown is purely plus polarized (i.e., ${\cal A}_\times=0$). This interesting property can be explained 
as follows.
{In the nonspinning case, and provided
\begin{equation}\label{eq:Rbarreal}
{\cal R}_{\chi=0}(\omega) = {\cal R}^*_{\chi=0}(-\omega^*) \,, 
\end{equation}
the transfer function satisfies the symmetry property
\begin{equation}\label{eq:kreal}
{\cal K}_{\chi=0}(\omega) = {\cal K}^*_{\chi=0}(-\omega^*)\,.
\end{equation}
The time domain 
echo waveforms are real (resp., imaginary) if the ringdown waveform is real (resp., imaginary).
In this case, the echo signal contains the same polarization of the BH ringdown and the two polarizations do not mix.
In particular, Eq.~\eqref{eq:Rbarreal} is satisfied when ${\cal R}$ is real.}

Remarkably, this property is broken in {the following} cases:
\begin{enumerate}
 \item when ${\cal R}$ is complex {and does not satisfy Eq.~\eqref{eq:Rbarreal}}
 , as in the second row of 
Fig.~\ref{fig:template_time};
 \item generically in the spinning case, even when ${\cal R}$ is real\footnote{In this case the transfer 
function 
satisfies an extended version of Eq.~\eqref{eq:kreal}, namely
\begin{equation}\label{eq:krealspin}
{\cal K}(\omega,m) = {\cal K}^*(-\omega^*,-m)\,
\end{equation}
which, however, does not prevent the mixing of the polarizations, due to the $m\to-m$ transformation.
} or when it satisfies 
Eq.~\eqref{eq:Rbarreal}.
\end{enumerate}
In either case \emph{mixing of the polarizations} occurs. For instance, if the BH ringdown is (say) a 
plus-polarized wave (${\cal A}_\times=0$), it might acquire a cross-polarization component upon reflection by the 
photon-sphere barrier (if $\chi\neq0$) or by the surface (if ${\cal R}$ is complex {and does not satisfy 
Eq.~\eqref{eq:Rbarreal}}). Therefore, even when the 
ringdown signal is linearly polarized (as when ${\cal A}_\times=0$, the case considered in 
Fig.~\ref{fig:template_time}), generically the final echo signal is not.

The mixing of polarizations can be used to explain the involved echo patter shown in some panels of 
Fig.~\ref{fig:template_time}. For example, for $\chi=0$ and ${\cal R}=e^{i\pi/3}$ each echo is multiplied by 
$e^{i\pi/3}$ relative to the previous one. Therefore, every three echoes the imaginary part of the signal (i.e., 
the cross polarization) is zero.

Another interesting consequence of the polarization mixing is the fact that the amplitude of subsequent echoes in 
each polarization does not decrease monotonically. This is evident, for example, in the panels of 
Fig.~\ref{fig:template_time} corresponding to $\chi=0.7$, ${\cal R}=1$ and $\chi=0$, ${\cal R}=e^{i\pi/3}$. 
However, it can be checked that the absolute value of the signal (related to the energy) decreases monotonically.

\subsubsection{Decay at late times and superradiant instability} \label{subsec:SR}
The involved behavior discussed above simplifies at very late times. In this case --~when the dominant frequency is 
roughly $\omega\approx \omega_R^{\rm ECO}\ll 1/M$ {(where $\omega_R^{\rm ECO}$ is the real part of the fundamental QNM 
of the ECO)}~-- the amplitude of the echoes always decreases 
as~\cite{Testa:2018bzd}
\begin{equation}
 |h_{\rm peaks}(t)|\propto |{\cal R}{\cal R}_{\rm BH}|^\frac{t}{2|x_0|} \,,\label{slope}
\end{equation}
where both ${\cal R}$ and ${\cal R}_{\rm BH}$ are evaluated at $\omega_R^{\rm ECO}\ll 1/M$. The 
above scaling agrees almost perfectly with our time-domain waveforms, especially at late times. 

More interestingly, Eq.~\eqref{slope} shows that the signal at late time should \emph{grow} when $|{\cal R}{\cal 
R}_{\rm BH}|>1$, i.e., when the combined action of reflection by the surface and by the BH barrier yields an 
amplification factor larger than unity~\cite{Maggio:2017ivp,Maggio:2018ivz}. When $|{\cal R}|\approx 1$, this condition 
requires 
\begin{equation}
 |{\cal R}_{\rm BH}|>1\,.
\end{equation}
From Eq.~\eqref{amplfactor}, it is easy to see that this occurs when
\begin{equation}
  \omega(\omega-m\Omega)<0\,, \label{SRcondition}
\end{equation}
i.e., when the condition for superradiance~\cite{zeldovich1,Teukolsky:1974yv} is satisfied~(see 
Ref.~\cite{Brito:2015oca} for an overview). Thus, we expect the signal to grow in time over a time scale given by the 
ergoregion instability~\cite{Friedman:1978wla,Cardoso:2007az,Cardoso:2008kj,Brito:2015oca,Moschidis:2016zjy,
Maggio:2017ivp,Maggio:2018ivz} of spinning horizonless ultracompact objects. Indeed, the QNM spectrum 
of the object contains unstable modes when 
$\omega_R<m\Omega$~\cite{Cardoso:2007az,Cardoso:2008kj,Maggio:2017ivp,Maggio:2018ivz}.
The instability time scale is always much 
longer than the dynamical time scale of the object (e.g., $\tau_{\rm instab}\gtrsim10^5 M$ for 
$\chi=0.5$~\cite{Maggio:2018ivz}).

When the signal grows in time due to the ergoregion instability the 
waveform $h(t)$ is a nonintegrable function, so its Fourier transform cannot be defined. For this reason the 
frequency-domain waveforms are valid up to $t\lesssim \tau_{\rm instab}$. Since the instability 
time scale is much longer than the echo delay time, the time interval of validity of our waveform still includes 
a large number of echoes. In particular, the ergoregion instability does not affect the first $N\sim|\log\delta/M|$ 
echoes~\cite{LRR}.

As discussed in Refs.~\cite{Maggio:2017ivp,Maggio:2018ivz}, this instability can be quenched if $|{\cal 
R}{\cal R}_{\rm BH}|<1$, which requires a partially absorbing ECO, $|{\cal R}|<1$ (see 
Refs.~\cite{Wang:2019rcf,Oshita:2019sat} for a specific model where the instability is absent).

\subsubsection{Energy of echo signal}
\begin{figure}[th]
\centering
\includegraphics[width=0.49\textwidth]{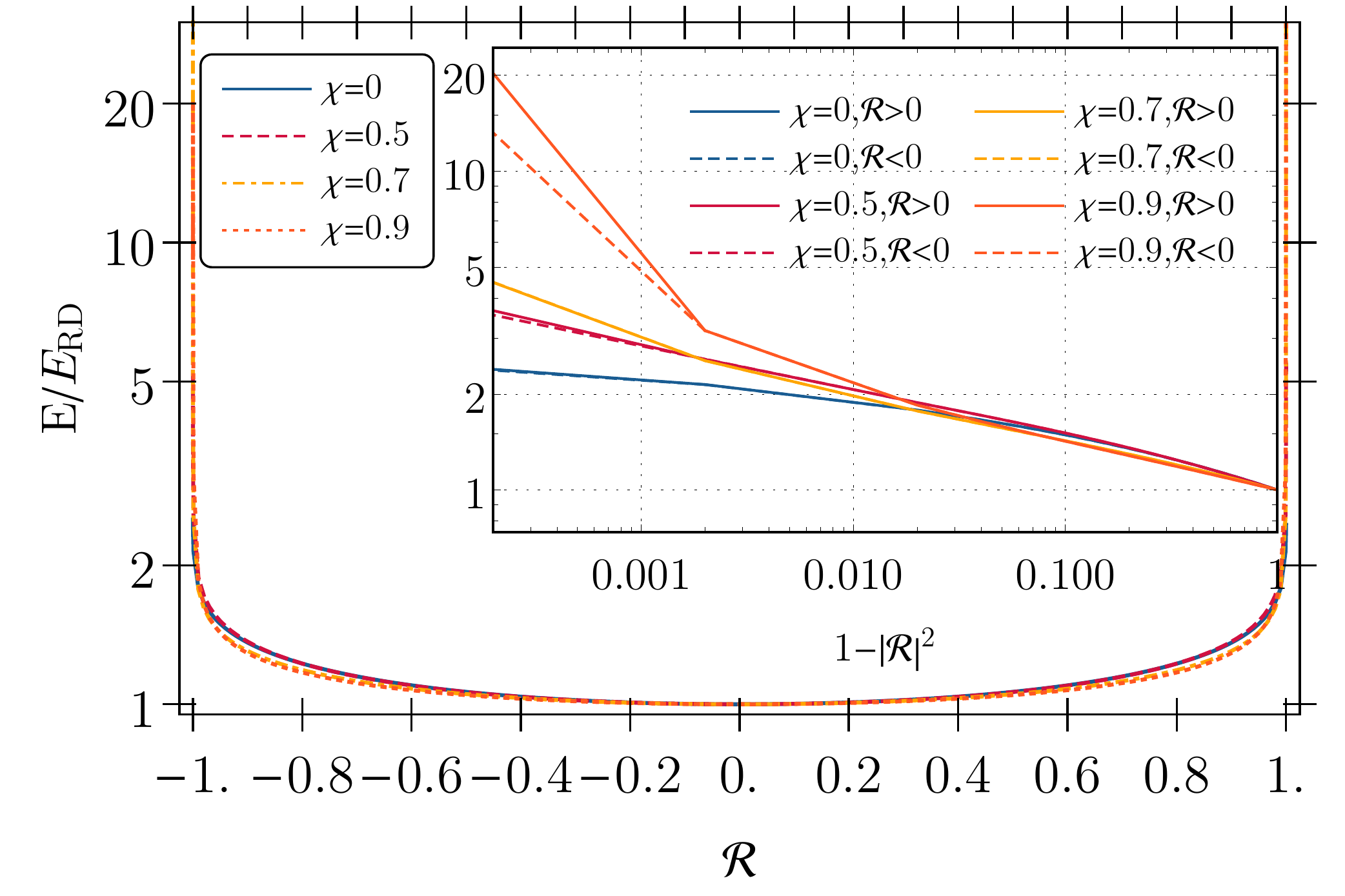}
\caption{Total energy contained in the ringdown$+$echo signal normalized by that of the ringdown alone as a function 
of ${\cal R}$ and for various values of the spin $\chi$. The total energy is much larger than the 
ringdown energy only when ${\cal R} \to 1$. We set $\delta/M=10^{-5}$ and considered only one 
ringdown polarization with $\phi_+=0$; the result is independent of $\delta$ in the $\delta\ll M$ limit.} 
\label{fig:EvsR}
\end{figure}

The energy contained in the ringdown$+$echo signal is shown in Fig.~\ref{fig:EvsR}, where we plot the quantity
\begin{equation}
 E\propto \int_{-\infty}^\infty d\omega\, \omega^2 |\hat Z^+|^2\,, \label{energy}
\end{equation}
normalized by the one 
corresponding to the ringdown alone, $E_{\rm RD} \equiv E({\cal R}=0)$, as a function 
of the reflectivity ${\cal R}$ and for several values of the spin $\chi$.
We use the prescription of Ref.~\cite{Flanagan:1997sx} to compute the ringdown energy, i.e. $\tilde{Z}^+_{\rm BH}$ is the 
frequency-domain full response obtained by using the Fourier transform of 
\begin{equation}
 Z_{\rm BH}^{+} (t)\sim \mathcal{A}_+\, \cos(\omega_R t+\phi_+)  e^{-|t|/\tau}\,.
\label{ZBHplusB}
\end{equation}
(Notice the absolute value of $t$ at variance with Eq.~\eqref{ZBHplus}.)
This prescription circumvents 
the problem associated with the Heaviside function in Eq.~\eqref{ZBHplus} that produces a spurious 
high-frequency behavior in the energy flux, leading to infinite energy in the 
ringdown signal. With the above prescription, the energy defined in 
Eq.~\eqref{energy} is finite and reduces to the result of 
Ref.~\cite{Flanagan:1997sx} for the BH ringdown 
when ${\cal R}=0$.

Because of reflection at the surface, the energy contained in the full signal for a fixed amplitude might be much 
larger than that of the ringdown itself. Overall, the normalized energy depends mildly on the spin, but 
much more strongly on ${\cal 
R}$: the energy contained in the echo part 
of the signal grows fast as $|{\cal R}|\to1$ (reaching a maximum value that depends on the spin and might become 
larger than the energy of the ringdown alone.
This is due to the resonances corresponding to the low-frequency QNMs of the 
ECO, that can be excited with large amplitude~\cite{Conklin:2017lwb} [see bottom panel of 
Fig.~\ref{fig:comparison_signal_FD}], and suggests that GW echoes 
might be detectable even when the ringdown is not if $|{\cal R}|\approx1$.
However, it is worth noticing that these low-frequency resonances are excited only at late times and therefore the 
first few echoes contain a small fraction of the total energy of the signal. When ${\cal R}$ is significantly 
smaller than unity subsequent echoes are suppressed (see third row in Fig.~\ref{fig:template_time}) and their total 
energy is modest compared to that of the ringdown.

Note also that when $|{\cal R}|\approx1$ the total energy is expected to diverge in the superradiant regime, due to 
the aforementioned ergoregion instability. This is not captured by the inverse Fourier transform $\hat Z^+(\omega)$, 
since the time-domain signal is non-integrable when $t\gtrsim \tau_{\rm instab}$.

\subsubsection{Frequency content of the signal} \label{spectrogram}
\begin{figure}[th]
 \centering
 \includegraphics[width=0.52\textwidth]{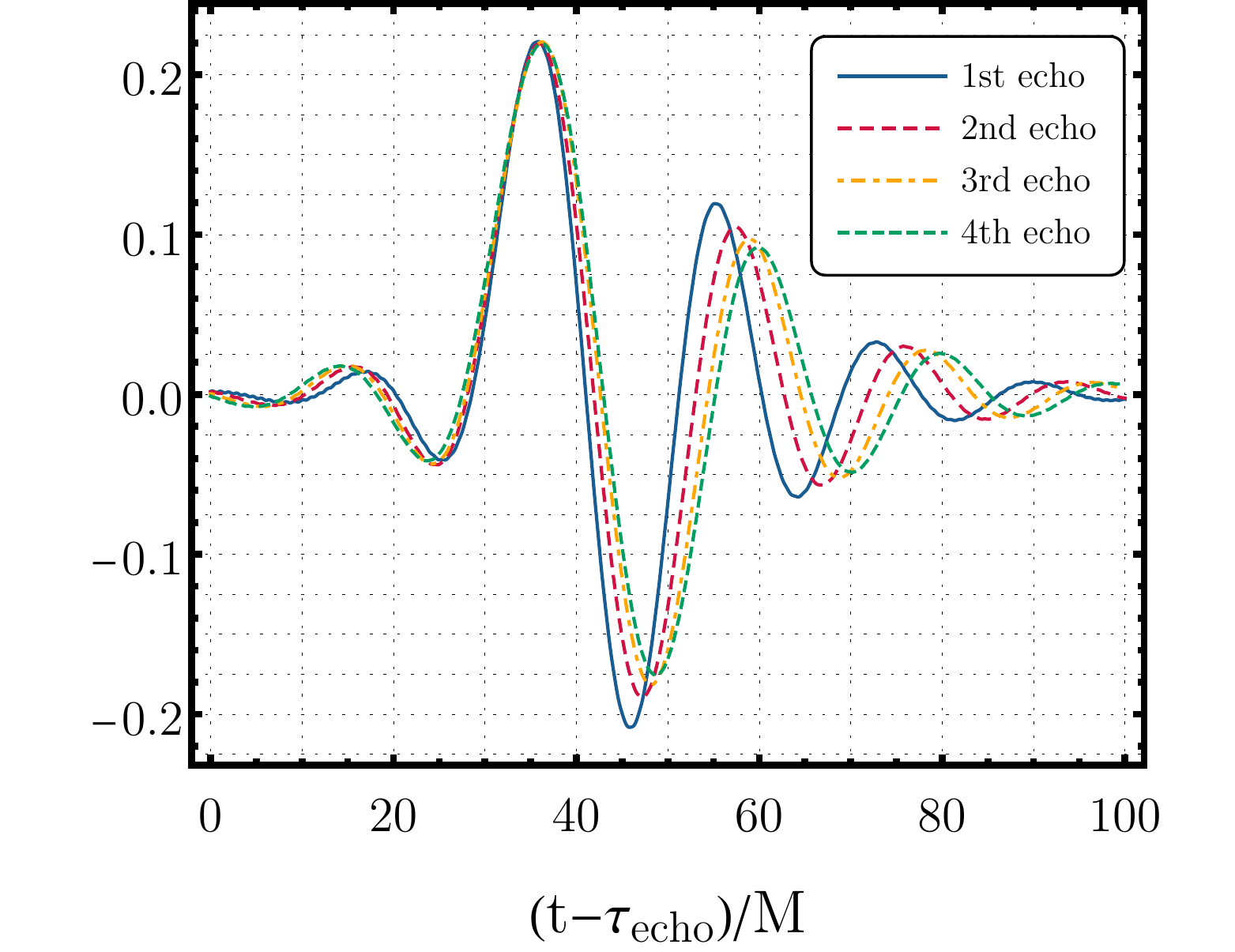}
 \caption{The first four echoes in the time-domain waveform for a model with ${\cal R}=1$, $\chi=0$, 
 $\delta/M=10^{-7}$. The waveform has been shifted in time and rescaled in amplitude so that the global 
 maxima of each echo are aligned. Note that each subsequent echo has a lower frequency content than the previous one.}
 \label{fig:freqcontent}
 \end{figure}

As previously discussed, the photon-sphere barrier acts as a high-pass filter as a consequence of which each echo has a lower 
frequency content than the previous one. This is confirmed by Fig.~\ref{fig:freqcontent}, where we display the first 
four echoes for ${\cal R}=1$, $\chi=0$, and $\delta/M=10^{-7}$, shifted in time and rescaled in amplitude so that 
their global maxima are aligned.

The frequency content of the total signal starts roughly at the BH QNM frequency, 
and slowly decreases in each subsequent echo until it is dominated by the low-frequency ECO QNMs at very late 
time. This also shows that a low-frequency approximation becomes increasingly more accurate at later times. In the 
example shown in Fig.~\ref{fig:freqcontent}, the frequencies of the first four echoes are approximately $M 
\omega\approx 0.34, 0.32, 0.3, 0.29$, whereas the real part of the fundamental BH QNM for $\chi=0$ is $M \omega_{R} 
\approx 0.37367$. Therefore, the frequency between the first and the fourth echo decreases by $\approx 17 \%$. 

Note that the case shown in Fig.~\ref{fig:freqcontent} is the one that provides the simplest echo patter ($\chi=0$, 
${\cal R}\in\Re$). The case $\chi\neq0$ or a complex choice of ${\cal R}$ would provide a much more involved 
patter and polarization mixing.

{Our results show that two qualitatively different situations can occur: 
\begin{itemize}
 \item[A)] the reflectivity ${\cal R}$ of the 
object is small enough so that the amplitude of subsequent echoes is suppressed. In this case most of the 
signal-to-noise ratio~(SNR) is contained in the first few echoes at frequency only slightly smaller than the 
fundamental BH QNM. 
 \item[B)] the reflectivity ${\cal R}$ is close to unity, so subsequent echoes are relevant and contribute 
significantly to the total SNR. In this case the frequency content becomes much smaller than the fundamental BH QNM.
\end{itemize}
Clearly our low-frequency approximation is expected to be accurate in case B) and less accurate in case A), especially 
for high spin where $M\omega_{\rm QNM}\sim 0.5$ or larger.
}

\subsubsection{On the phase of the reflectivity coefficients} \label{sec:phase}
It is worth remarking that there exist several definitions of the radial function describing the perturbations of a 
Kerr metric; these are all related to each other by a linear transformation similar to Eq.~\eqref{DetweilerX}. The BH 
reflection coefficients that can be defined for each function differ by a phase, while the quantity $|{\cal R}_{\rm 
BH}|^2$ (related to the energy damping/amplification) is invariant~\cite{Chandra}.

The transfer function in Eq.~\eqref{transfer} contains both the absolute value and the phase of ${\cal R}_{\rm BH}$. 
Therefore, one might wonder whether this ambiguity in the phase could affect the ECO response. For a given 
model, it should be noted that the reflectivity coefficient at the surface, ${\cal R}$, is also affected by the same 
phase ambiguity, in accordance with the perturbation variable chosen to describe the problem. Since the transfer 
function depends only on the combinations ${\cal R}{\cal R}_{\rm BH}$ and ${\cal R}{\cal T}_{\rm BH}$, the phase 
ambiguity in ${\cal R}$ cancels out with that in ${\cal R}_{\rm BH}$ and ${\cal T}_{\rm BH}$ in Eq.~\eqref{transfer}. 
This ensures that the transfer function is invariant under the choice of the radial perturbation function, as expected 
for any measurable quantity. For example, at small frequencies the BH reflection coefficient derived from the 
asymptotics of the Regge-Wheeler function at $x\to-\infty$ has a phase difference of $\pi$ compared to the BH 
reflection coefficient computed from the Detweiler function for $\chi=0$. Consistently, the reflectivity coefficient 
associated to the former differs by a phase $\pi$ with respect to the reflectivity coefficient associated to the latter, 
i.e., if $\bar{\mathcal{R}} = 1$ for Regge-Wheeler then $\bar{\mathcal{R}} = -1$ for Detweiler in the same model, and 
viceversa.

Therefore, it is natural for ${\cal R}$ to have a nontrivial (and generically frequency- and spin- dependent) phase 
term, whose expression depends on the formulation of the problem. Obviously, all choices of the radial wavefunctions are 
equivalent but --~for the same ECO model~-- the complex reflection coefficient ${\cal R}$ should generically be 
different for each of them.
To the best of our knowledge, this point was neglected in actual matched-filtered searches for echoes, which so far 
considered ${\cal R}$ (and also ${\cal R}_{\rm BH}$) to be real.

This fact is particularly important in light of what previously discussed for the mixing of the polarizations. As shown 
in the second row of Fig.~\ref{fig:template_time}, a phase in ${\cal R}$ introduces a mixing of polarizations for 
any spin, which results in a more complex shape of the echoes in the individual polarizations of the signal.

Since the phase of ${\cal R}$ depends on the specific ECO model, in the analysis of Sec.~\ref{sec:bounds} we will 
parametrize the reflectivity in a model-agnostic way as ${\cal R}=|{\cal R}|e^{i\phi}$. In principle, both the 
absolute value and the phase are generically frequency dependent but for 
simplicity we choose them to be constants or, equivalently, we take the leading-order and low-frequency limit of 
these quantities. Hence we parametrize our template by $|{\cal R}|$ and $\phi$, different choices of which 
correspond to different 
models.

\subsubsection{BH QNMs vs ECO QNMs} \label{sec:poles}
It is worth considering the inverse-Fourier transform of Eq.~\eqref{signalomega} (i.e., Eq.~\eqref{inverseFT})
and deform the frequency integral in the complex frequency plane. When ${\cal R}=0$ (i.e., standard BH ringdown) this 
procedure yields three contributions~\cite{Leaver:1986gd,Berti:2009kk}: (i) the high-frequency 
arcs that govern the prompt response, (ii) a sum-over-residues at the poles of 
the complex frequency plane (defined by $W_{\rm BH}=0=B_{\rm out}$), which correspond to 
the QNMs and dominate the signal at intermediate times, and (iii) a branch cut on 
the negative half of the imaginary axis, giving rise to late-time tails due to 
backscattering off the background curvature.

When ${\cal R}\neq0$, the pole structure is more involved. The extension of the integral in Eq.~\eqref{inverseFT} 
to the complex plane contains two types of complex poles: (i)~those associated with $\tilde Z_{\rm 
BH}^+(\omega)$ ($\sim 1/W_{\rm BH}\sim 1/B_{\rm out}$) and with ${\cal K}\tilde Z_{\rm 
BH}^-(\omega)$ ($\sim \mathcal{T}_{\rm BH}/W_{\rm BH}\sim 1/B_{\rm out}^2$) which are the standard BH QNMs (but that do not 
appear in the ECO QNM spectrum~\cite{Cardoso:2016rao}), and (ii)~those associated with the poles of the transfer 
function ${\cal K}$ (i.e. when ${\cal R}_{\rm BH}={e^{2 i k x_0}}/{\cal R}$), which correspond to the ECO QNMs.

The late-time signal in the post-merger is dominated by the second type of poles, since the 
latter have a longer damping time and survive longer. The prompt 
ringdown is dominated by the first type of poles, i.e., by the dominant QNMs of 
the corresponding BH spacetime~\cite{Cardoso:2016rao}.
Finally, the intermediate region between prompt ringdown and late-time ECO QNM ringing depends on the other parts of 
the contour integral on the complex plane. As such, they are more complicated to model, since they do not depend on the QNMs 
alone and might also depend on the source, as in the standard BH case.

\section{Projected constraints on ECOs} \label{sec:bounds}
%
In this section we use the template derived in Sec.~\ref{sec:template} for a preliminary error estimation of 
the ECO properties using current and future GW detectors.

The ringdown$+$echo signal displays sharp peaks which originate from the resonances of the transfer 
function ${\cal K}$ and correspond to the long-lived QNMs of the ECO~\cite{Maggio:2018ivz}. The 
relative amplitude of each resonance in the signal depends on the source and the dominant modes are not necessarily the 
fundamental harmonics~\cite{Mark:2017dnq,Bueno:2017hyj}.
We stress that the amplitude of the echo signal depends strongly on the value of ${\cal R}$, especially when 
$|{\cal R}|\approx1$. This suggests that the 
detectability of (or the constraints on) the echoes strongly depends on ${\cal 
R}$ and would be much more feasible when $|{\cal R}|\approx 1$. Below we 
quantify this expectation using a Fisher matrix technique, which is accurate at large SNR (see, e.g., 
Ref.~\cite{Vallisneri:2007ev}). 
This is performed as in Ref.~\cite{Testa:2018bzd}, but by 
including the spin of the object consistently and allowing for a 
complex reflection coefficient, ${\cal R}=|{\cal R}|e^{i\phi}$. 
%

The Fisher information matrix $\Gamma$ of a template $\tilde h(f)$ for a detector with 
noise spectral density $S_n(f)$ reads as
\begin{equation}\label{fisher}
\Gamma_{i j} = \langle \partial_i \tilde h |  \partial_j \tilde h\rangle \,,
\end{equation}
where $i,j=1,...,N$, with $N$ being the number of parameters in the template.
The SNR $\rho$ is defined such as $\rho^2 = \langle \tilde h | \tilde h\rangle$.
%
%
The covariance matrix, $\Sigma_{ij}$, of the errors on the template's parameters is the inverse of $\Gamma_{ij}$ 
and $\sigma_{i}=\sqrt{\Sigma_{ii}}$ (no summation) gives the statistical error associated with the measurement of 
$i$-th parameter.

%
We computed numerically the Fisher 
matrix~\eqref{fisher} with our template $\tilde h(f)\equiv \tilde Z^+(f)$ using the 
sensitivity curves of aLIGO with the design-sensitivity \texttt{ZERO\_DET\_high\_P}~\cite{zerodet} and two 
configurations for the third-generation~(3G) instruments: Cosmic~Explorer in 
the narrow band variant~\cite{Evans:2016mbw,Essick:2017wyl}, and Einstein Telescope 
in its \texttt{ET-D} configuration~\cite{Hild:2010id}. We also consider the LISA's noise 
spectral density proposed in Ref.~\cite{LISA}.
We focus on the most relevant case of gravitational perturbations with $l=m=2$ and consider $M=30\,M_\odot$ 
($M=10^6\,M_\odot$) for ground- (space-) based detectors.

As previously discussed, the most generic BH ringdown template 
contains $7$ parameters (mass, spin, two phases, two amplitudes and starting 
time). For simplicity, we reduce it to a linearly-polarized ringdown. In particular, we do not include ${\cal 
A}_\times$ and $\phi_\times$ in the parameters and inject ${\cal A}_\times=0$. This implies that we have $5$ 
standard-ringdown parameters in our analysis. 

The template also depends on two ECO quantities (the frequency-dependent reflection 
coefficient ${\cal R}(\omega)$ and the parameter $\delta$) which 
fully characterize the model. The parameter $\delta$ is directly related to 
physical quantities, in particular, the compactness of the ECO or (equivalently) the redshift at 
the surface.
We parametrize the reflectivity coefficient as
\begin{equation}
 {\cal R}(\omega)={|{\cal R}|e^{i\phi}} \,,
\end{equation}
where $|{\cal R}|$ and $\phi$ are assumed to be frequency independent for simplicity and we remark that 
$x_0=x_0(\delta)$ (see Eq.~\eqref{proper}). This yields 
three ECO parameters: $\delta$, $|{\cal R}|$, and $\phi$.

We consider two cases: (i)~a conservative case in which we extract the errors on all the $5+3$ 
parameters in a Fisher matrix framework and (ii) a more optimistic case in which we assume that the standard-ringdown 
parameters can be independently and reliably measured through the prompt ringdown, so that we are left with the 
measurements errors on the $3$ ECO parameters.

\subsection{Conservative case: $5+3$ parameters}
 \begin{figure*}[th]
 \centering
 \includegraphics[width=0.32\textwidth]{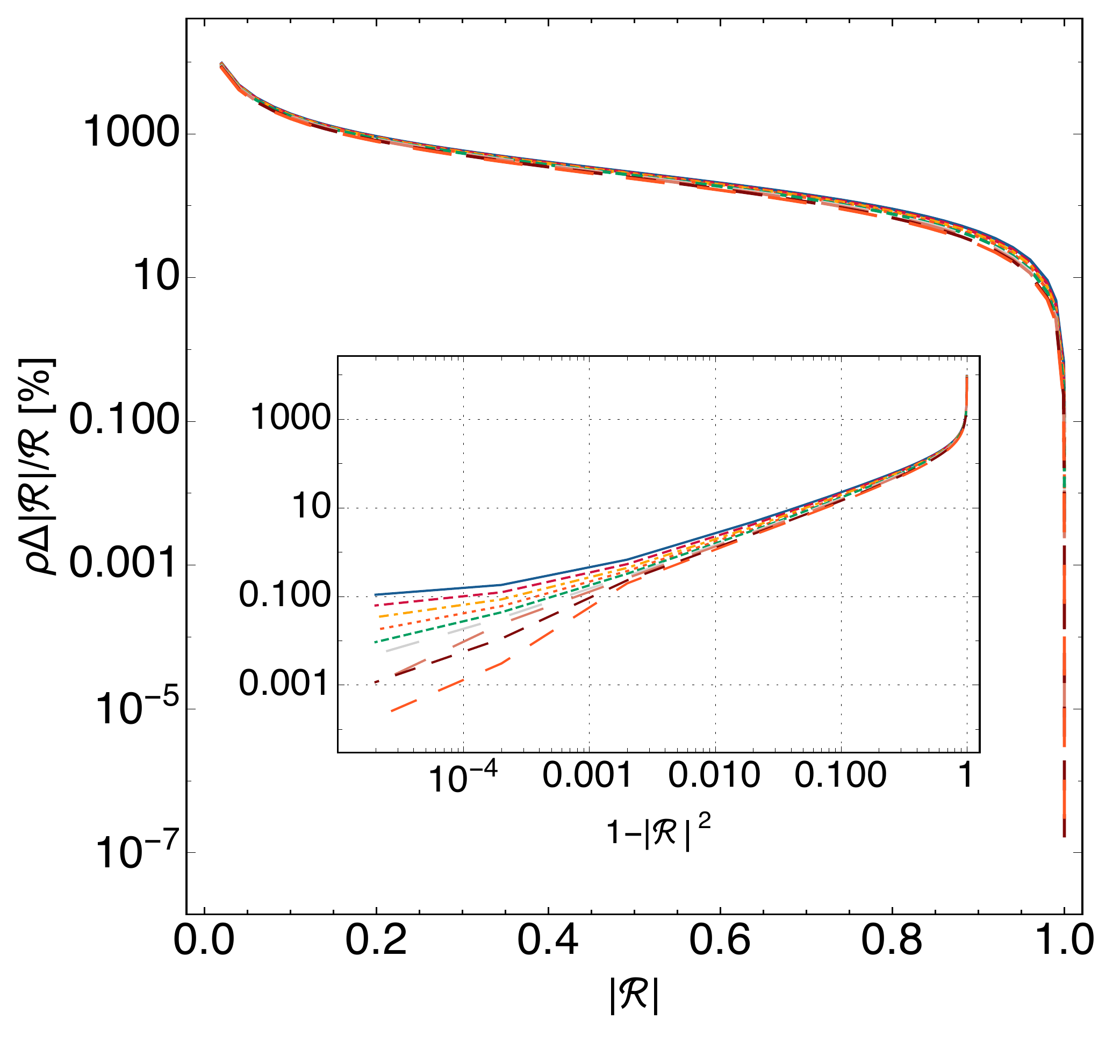}
 \includegraphics[width=0.32\textwidth]{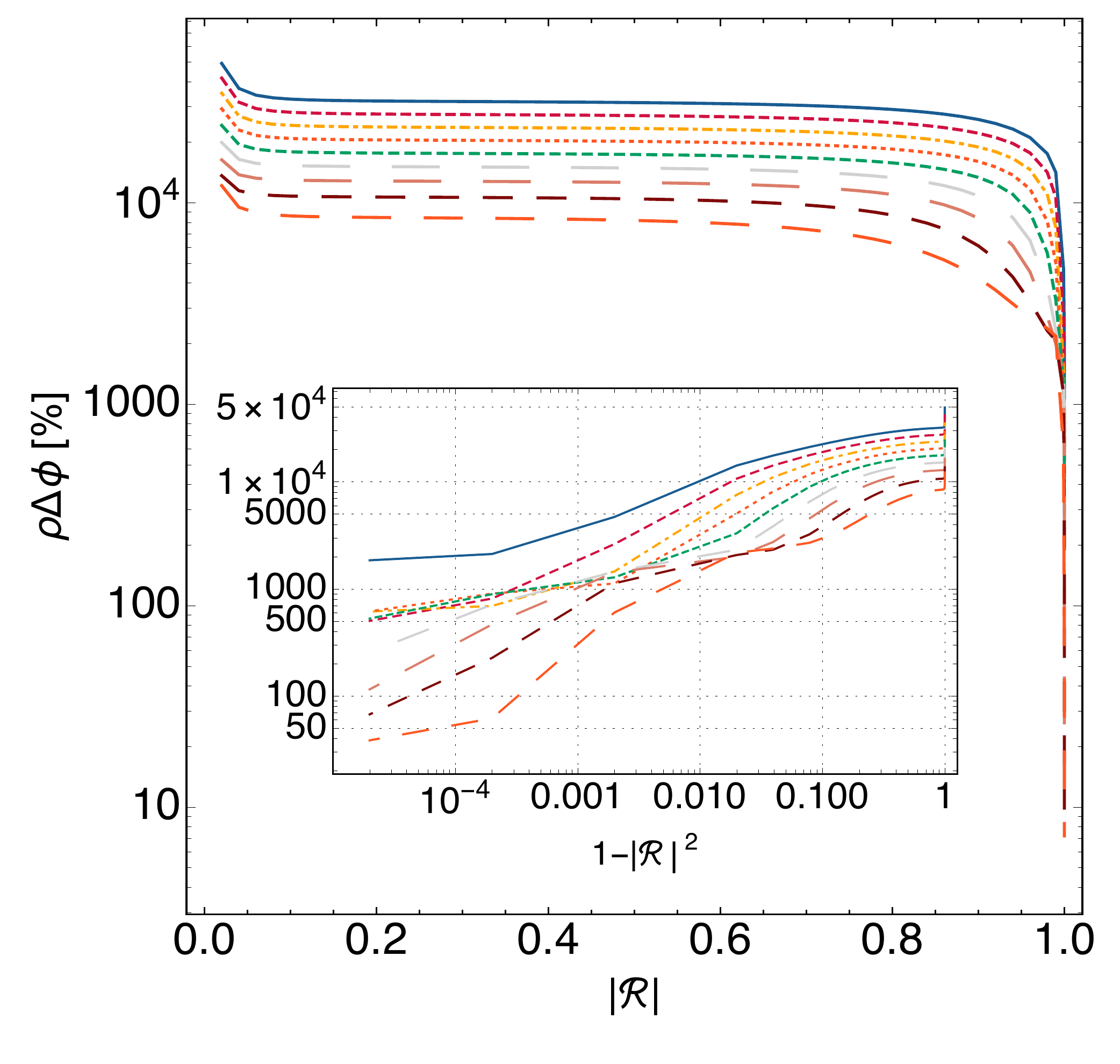}
 \includegraphics[width=0.32\textwidth]{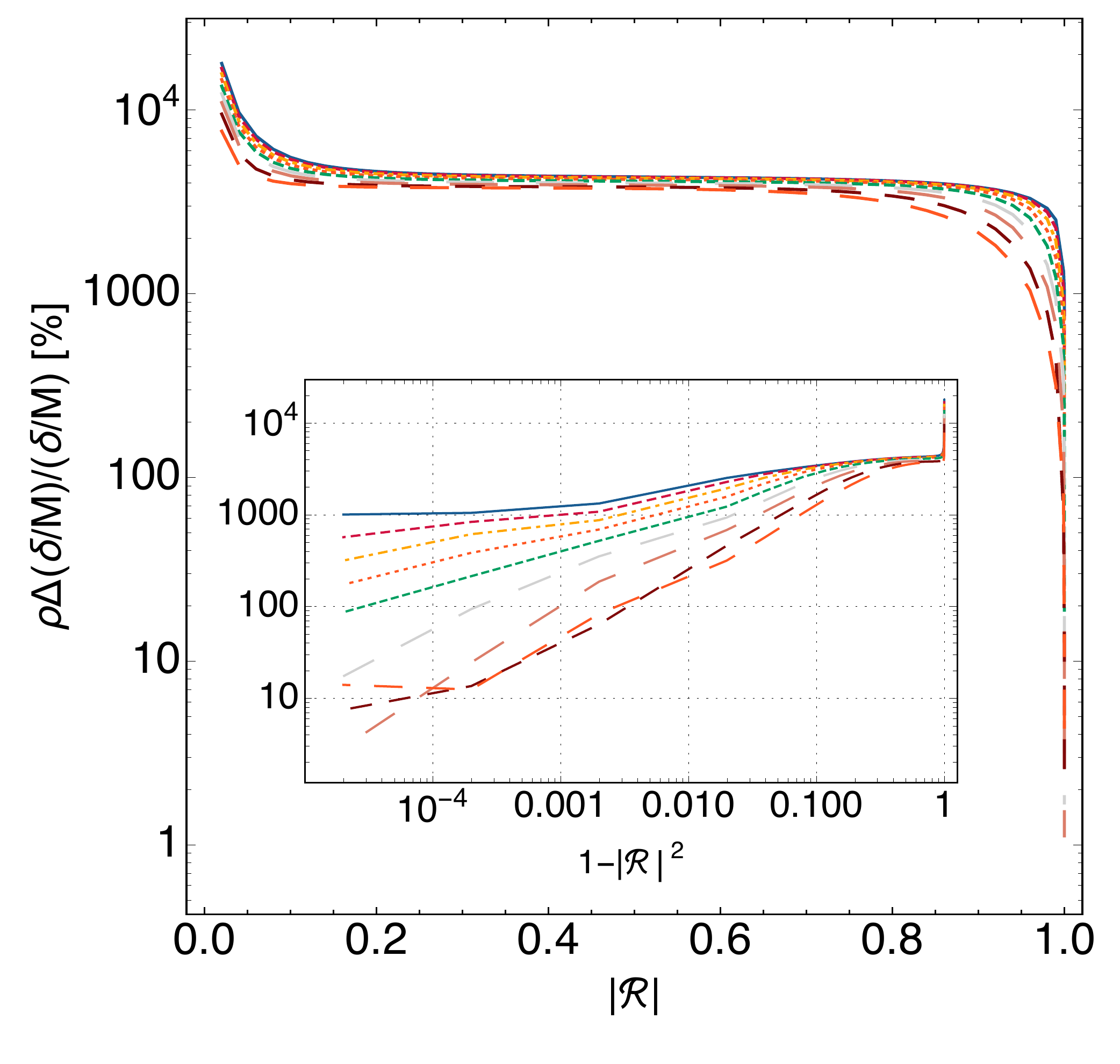}
 \caption{Left panel: relative (percentage) error on the reflection coefficient, 
 $\Delta |{\cal R}| / |{\cal R}|$ multiplied by the SNR, as a function of $|{\cal R}|$ for 
different values of injected spin. The inset shows the same quantity as a function of $1-|{\cal R}|^2$ in a 
 logarithmic scale. From top to bottom: $\chi=(0.9,0.8,0.7,0.6,0.5,0.4,0.3,0.2,0.1)$.
 Middle panel: same but for the absolute (percentage) error on the phase $\phi$ of ${\cal R}$, i.e. 
$\rho\Delta\phi$.
 Right panel: same as 
 in the left panel but for the compactness parameter, $\delta$, i.e.
 $\Delta (\delta/M)/(\delta/M)$. 
 We assume $\delta=10^{-7} M$ but the errors are independent 
 of $\delta$ when $\delta/M\ll1$~\cite{Testa:2018bzd}. We set $\phi=0$ for the phase of ${\cal R}$ (i.e. we 
 consider a real and positive ${\cal R}$, but other choices give very similar results.} 
 \label{fig:errors}
 \end{figure*}
%
Our main results for the statistical errors on the ECO parameters are 
shown in Fig.~\ref{fig:errors}. In the large SNR limit, the errors scale as 
$1/\rho$ so we present the quantity $\rho\Delta |{\cal R}| / |{\cal R}|$ (left panel), $\rho\Delta \phi 
$ 
(middle panel), and $\rho\Delta (\delta/M)/(\delta/M)$ (right panel) for several values of the spin.
We find that the main qualitative features already discussed in Ref.~\cite{Testa:2018bzd} do not depend significantly 
on the inclusion of the spin in the template. In particular, for fixed SNR the relative errors are almost independent 
of the specific sensitivity curve of the detector, at least for signals located near each minimum of the sensitivity 
curve, as those adopted in Fig.~\ref{fig:errors}. In Fig.~\ref{fig:errors} we adopted the LISA curve~\cite{LISA} 
but other detectors give very similar results for the errors normalized by the SNR. 

Furthermore, the statistical errors are almost independent 
of $\delta$ when $\delta/M\ll1$, whereas they strongly depend on the reflection 
coefficient ${\cal R}$. The reason for this can be again traced back to the presence of resonances as ${\cal 
R}\to1$.
This feature confirms that it should be relatively straightforward 
to rule out or detect models with $|{\cal R}|\approx 1$, whereas it is 
increasingly more difficult to constrain models with smaller values of $|{\cal 
R}|$.

We also note that the value of the spin of the remnant affects the errors on $|{\cal R}|$ only mildly, whereas 
it 
has a stronger impact on the phase of ${\cal R}$ (probably due to the aforementioned mixing of the 
polarizations) 
and a moderate impact on the errors on $\delta$.

Overall, the specific value of $\phi$ does not affect the errors significantly, although it is important to include it 
as an independent parameter in order not to underestimate the errors.

\begin{figure}[th]
 \centering
 \includegraphics[width=0.49\textwidth]{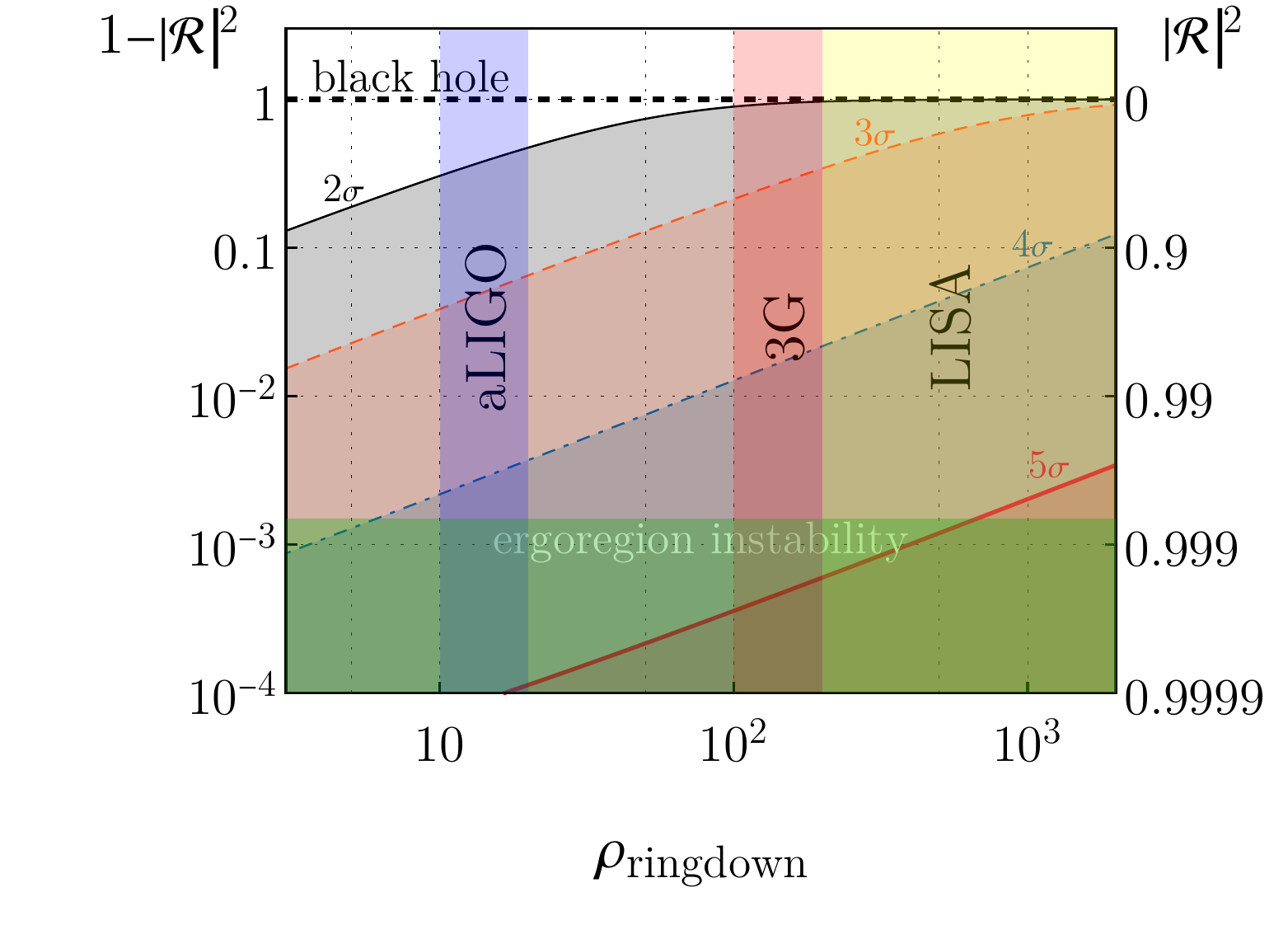}
 \caption{Projected exclusion plot for the ECO reflectivity ${\cal R}$ as a 
 function of the SNR in the ringdown phase. The shaded areas represent regions that can be excluded at a 
 given confidence level ($2\sigma$, $3\sigma$, $4\sigma$, $5\sigma$). 
 Vertical bands are typical SNR achievable by aLIGO/Virgo, 3G, and LISA in the ringdown phase, whereas the horizontal 
 band is the region excluded by the ergoregion instability~\cite{Maggio:2017ivp,Maggio:2018ivz}. We assumed $\chi=0.7$ 
for the spin of the merger remnant, the result depends only mildly on the spin.} \label{fig:RvsSNR}
 \end{figure}

Next, we calculate the SNR necessary to discriminate a 
partially-absorbing ECO from a BH on the basis of a measurement of ${\cal R}$ 
at some confidence level~\cite{Testa:2018bzd}. Clearly, if $\Delta {\cal R}/{\cal R}>100\%$, any 
measurement would be compatible with the BH case (${\cal R}=0$). On the other 
hand, relative errors $\Delta {\cal R}/{\cal R}<(4.5,0.27,0.007,0.00006)\%$ suggest that it is possible
to detect or rule out a given model at $(2,3,4,5)\sigma$ confidence level, 
respectively.
The result of this analysis is shown in Fig.~\ref{fig:RvsSNR}, 
where we present the exclusion plot for the parameter ${\cal R}$ as a function 
of the SNR in the ringdown phase only, $\rho_{\rm RD}$. 
Shaded areas represent regions which can be 
excluded at some given confidence level. Obviously, larger SNRs would allow to 
probe values of ${\cal R}$ close to the BH limit, ${\cal R}\approx 0$.
The extent of the constraints strongly depends on the confidence 
level. For example, ${\rm SNR\approx100}$ in the ringdown would allow to distinguish 
ECOs with $|{\cal R}|^2\gtrsim 0.1$ from BHs at $2\sigma$ confidence level, but a 
$3\sigma$ detection would be possible only if $|{\cal R}|^2\gtrsim0.8$. The reason for this is 
again related to the strong dependence of the echo signal on ${\cal R}$ .
Note that Fig.~\ref{fig:RvsSNR} is very similar to that computed in Ref.~\cite{Testa:2018bzd}, showing that including 
the spin and a phase term for ${\cal R}$ does not affect the final result significantly.

%
%

\subsection{Optimistic case: $3$ ECO parameters}
Let us now assume that the standard ringdown parameters (mass, spin, phases, amplitudes, and starting 
time) can be independently measured through the prompt ringdown signal, which is identical for BHs and ECOs if 
$\delta/M\ll1$~\cite{Cardoso:2016rao}. In such case the remaining three ECO parameters ($|{\cal R}|$, $\phi$, 
and 
$\delta$) can be measured a posteriori, assuming the standard ringdown parameters are known.

\begin{figure*}[th]
 \centering
 \includegraphics[width=0.32\textwidth]{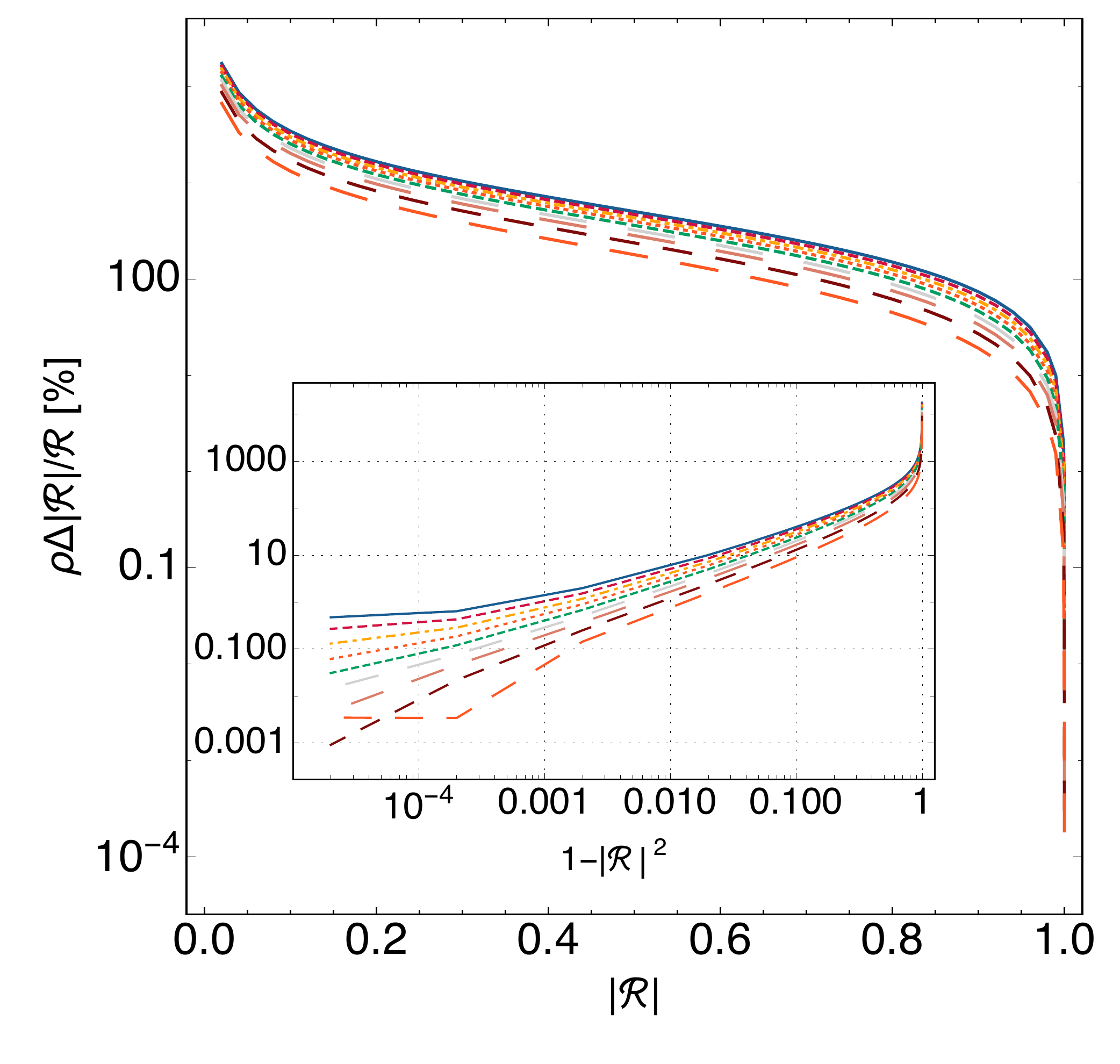}
 \includegraphics[width=0.32\textwidth]{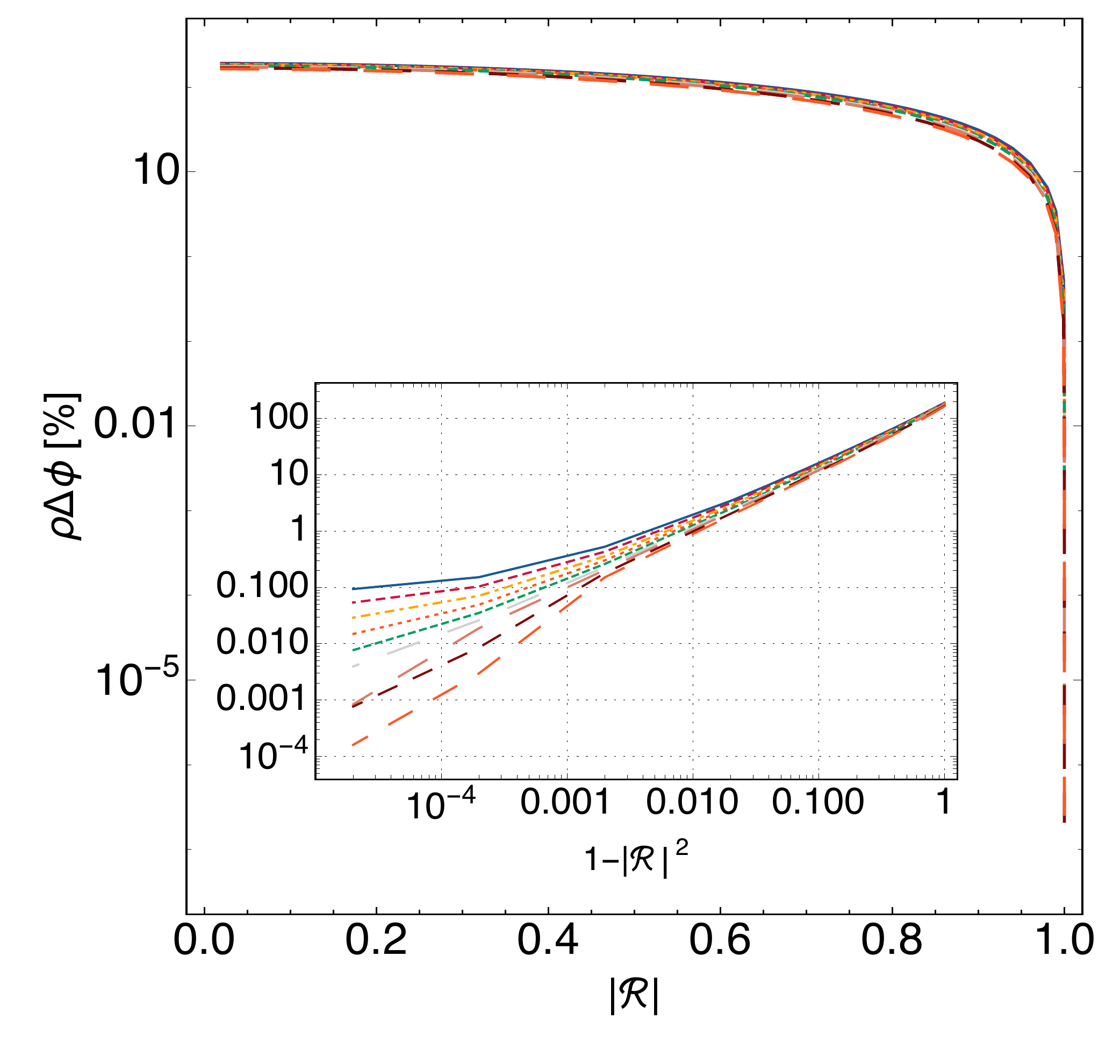}
 \includegraphics[width=0.32\textwidth]{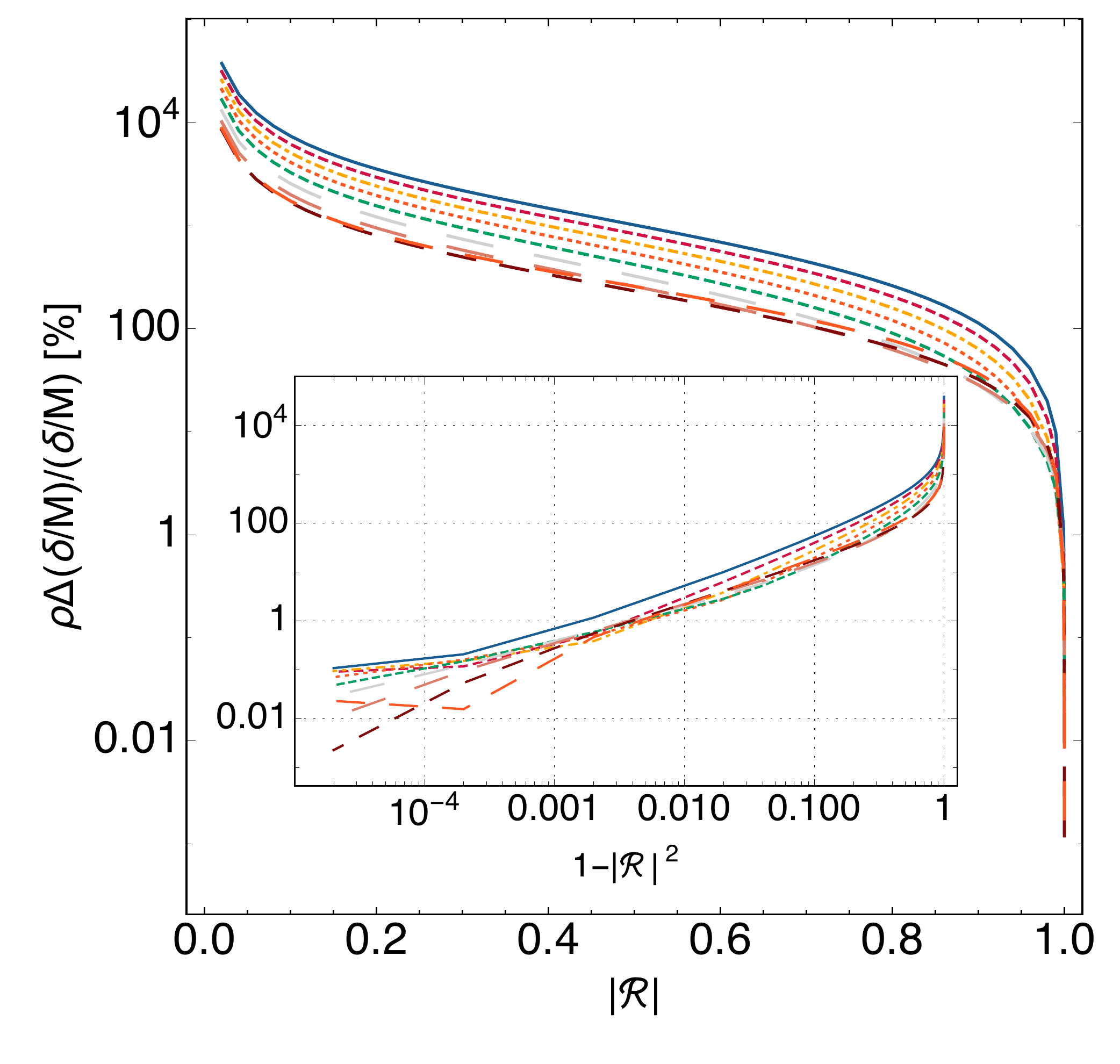}
 \caption{Same as in Fig.~\ref{fig:errors} but including only the three ECO parameters ($|{\cal R}|$, $\phi$, and 
 $\delta$) in the Fisher analysis.} 
 \label{fig:errors3param}
 \end{figure*}

A representative example for this optimistic scenario is shown in Fig.~\ref{fig:errors3param}. As expected, the errors 
are significantly smaller, especially those on the phase $\phi$ of 
the reflectivity. The errors on ${\cal R}$ are only mildly affected, and the projected constraints on ${\cal R}$ at 
different confidence levels are similar to those shown in Fig.~\ref{fig:RvsSNR}.
Nonetheless, we expect this strategy to be much more effective for actual searches.

%

\section{Discussion}~\label{sec:Discussion}
%
We have presented an analytical template that describes the ringdown and 
subsequent echo signal of a spinning, ultracompact, Kerr-like horizonless object.
This template depends on the physical parameters of the remnant: namely, the mass, the spin, the compactness and the
reflection coefficient ${\cal R}$ at its surface.
The analytical approximation is valid at low frequencies, where most of the SNR of an echo signal is accumulated 
{in the case $|{\cal R}|\sim 1$}. Our template becomes increasingly accurate at later times as the frequency 
content of the echo decreases. 

The features of the signal are related to the physical properties of the ECO model. The time-domain 
waveform contains all features previously reported for the echo signal, namely amplitude and frequency modulation and 
possible phase inversion of each echo relative to the previous one, depending on the reflective boundary conditions. 
Furthermore, the presence of the spin and of a generically complex reflectivity introduce qualitatively different 
effects, most notably the amplitude and frequency modulation is more involved (also) due to mixing of the two 
polarizations. For (almost) perfectly-absorbing spinning ECOs, the perturbations can grow at late times due to 
superradiance and the ergoregion instability. However, even for highly-spinning remnants, this instability occurs on a 
time scale which is much longer than the echo delay time, and likely plays a negligible role in actual searches for 
echoes (see however Ref.~\cite{Barausse:2018vdb} for a discussion of the stochastic background produced by this 
instability). The instability is quenched for partially-reflecting 
objects~\cite{Maggio:2017ivp,Maggio:2018ivz,Wang:2019rcf,Oshita:2019sat}.

The amplitude of subsequent echoes depends strongly on the reflectivity ${\cal R}$. When $|{\cal R}|\approx 
1$ the echo signal can have energy significantly larger than those of the ordinary BH ringdown. This suggests that GW 
echoes in certain models might be detectable even when the ringdown is not. Likewise, ruling out models with $|{\cal 
R}|\approx 1$ is significantly easier than for smaller values of the reflectivity.
 
We have also highlighted the importance of including a model-dependent phase term in the reflection coefficient; this
phase also depends on the radial perturbation variable used in the perturbation equation. To the best of our knowledge 
this issue has been so far neglected in previous analyses (but see Ref.~\cite{UchikataEchoes} for a recent discussion).
We showed that a complex reflectivity at the surface (or, generically, the spin of the remnant) introduce mixing among 
the two polarizations, drastically modifying the shape of the echoes.

Using a Fisher analysis, we have then estimated the statistical errors on the 
template parameters for a post-merger GW detection with current and future GW
interferometers. Our analysis suggests that ECO models 
with $|{\cal R}|^2\approx 1$ can be detected or ruled out with aLIGO/Virgo 
(for events with $\rho_{\rm ringdown}\gtrsim 8$) at $5\sigma$ 
confidence level. These events might also allow us to probe values of 
the reflectivity as small as $|{\cal R}|^2\approx 0.8$ at $\approx 2\sigma$ 
confidence level.

ECOs with $|{\cal R}|=1$ are already ruled out by the ergoregion 
instability~\cite{Cardoso:2008kj,Maggio:2017ivp} and by the 
absence of GW stochastic background in LIGO O1 run~\cite{Barausse:2018vdb}. 
Excluding/detecting echoes for models with smaller values of the reflectivity 
(for which the ergoregion instability is absent~\cite{Maggio:2017ivp,Maggio:2018ivz})
requires SNRs in the post-merger phase of ${\cal O}(100)$. This will be achievable only 
with 3G detectors (ET and Cosmic Explorer) and with the 
space-based mission LISA.
Our preliminary analysis confirms that very stringent constraints on (or 
detection of) ultracompact horizonless objects can be obtained with current (and 
especially future) interferometers.

Several interesting extensions of this work are left for the future. In a follow-up paper we plan to adopt the template 
developed here in a matched-filtered search for GW echoes using LIGO/Virgo public data and for a Bayesian parameter 
estimation. This can be done for a generic reflectivity coefficient ${\cal R}$, or for specific models, such as those 
motivated by effective field theory arguments~\cite{Burgess:2018pmm} and the model recently proposed in 
Refs.~\cite{Wang:2019rcf,Oshita:2019sat} for the Boltzmann reflectivity of quantum BHs.

An important open problem is to compare the echo template (obtained within perturbation theory) with the post-merger 
signal of an ECO coalescence producing an echoing merger. Unfortunately, numerical simulations of these systems are 
currently unavailable and so are inspiral-merger-ringdown waveforms for these models. Assessing the reliability of the 
analytical template and the importance of nonlinearities will 
require a comparison between analytical and numerical waveforms, following a path similar to what done in 
the past for the matching of standard BH ringdown templates with numerical-relativity 
waveforms (see, e.g., Ref.~\cite{Buonanno:2006ui}).

A more technical extension deals with the modeling of the signal beyond the low-frequency approximation. The 
characteristic frequency of the echo signal is always smaller than the corresponding BH ringdown frequency. We expect 
our template to be robust to the prescription for transition to high frequencies.
Nevertheless, it might be interesting to develop a high-frequency analytical approximation of the BH reflection and 
transmission coefficients to be matched smoothly with a low-frequency approximation. By performing the low-frequency and 
high-frequency expansions beyond the leading order it might be possible to obtain a better analytical approximation of 
the transfer function at all frequencies.

\begin{acknowledgments}
We are indebted to Emanuele Berti, Gregorio Carullo, Walter Del Pozzo, Antoine Klein, {Simone Mastrogiovanni} and John 
Veitch for interesting discussions and correspondence, and to Xin Shuo for highlighting a typo in 
Eq.~\eqref{eq:bhtemplateINF}.
We acknowledge support provided under the European Union's H2020 ERC, Starting 
Grant agreement no.~DarkGRA--757480 and by the Amaldi Research Center funded by the MIUR program "Dipartimento di 
Eccellenza" (CUP: B81I18001170001). AT is also grateful for the support provided by the Walter Burke Institute for 
Theoretical Physics.
\end{acknowledgments}
%

\appendix
%
\section{Low-frequency solution of Teukolsky equation} \label{app:lowfreq}
%
In this appendix we derive an analytical solution for the reflection coefficient of a BH for 
gravitational 
perturbations in the small-frequency regime through a matched asymptotic expansion. The technique is detailed 
in Ref.~\cite{Maggio:2018ivz}.

For generic spin-$s$ perturbations, Teukolsky's equations are~\cite{Teukolsky:1972my,Teukolsky:1973ha,Teukolsky:1974yv}
\begin{eqnarray}
&&\Delta^{-s} \frac{d}{dr}\left(\Delta^{s+1} \frac{d_{s}R_{lm}}{dr}\right)\nonumber\\
&+& \left[\frac{K^{2}-2 i s (r-M) K}{\Delta}+4 i s \omega r -\lambda_s\right] \ _{s}R_{l 
m}=0\,,\label{wave_eq} \quad \\
&&\left[\left(1-y^2\right)~_{s}S_{l m,y}\right]_{,y}+ \bigg[(a\omega y)^2-2a\omega s y + s 
\nonumber\\
&+&~_{s}A_{lm}-\frac{(m+sy)^2}{1-y^2}\bigg]~_{s}S_{l m}=0\,, \label{angular}
\end{eqnarray}
where $~_{s}S_{l m}(\theta)e^{im\phi}$ are spin-weighted spheroidal
harmonics, $y\equiv\cos\theta$, and the separation
constants
$\lambda$ and $~_{s}A_{l m}$ are related by $\lambda_s \equiv ~_{s}A_{l
m}+a^2\omega^2-2am\omega$.

In the region near the surface of the ECO, the radial wave equation~\eqref{wave_eq} for $M \omega \ll 1$ reduces to \cite{Starobinskij2}
\begin{eqnarray}
&& \left[z(z+1)\right]^{1-s} \partial_z \left\{ \left[z(z+1)\right]^{s+1} \partial_z R_s \right\} \nn \\
&& \quad +\left[ Q^2 + i Q s (1+2z) - (l-s)(l+s+1) z (z+1)\right] R_s = 0 \,, \nn \\ \label{wave_eq_near_hor}
\end{eqnarray}
where $z=(r-r_+)/(r_+ - r_-)$ and $R_s \equiv ~_{s}R_{lm}$ for brevity. The general solution of 
Eq.~\eqref{wave_eq_near_hor} is a linear combination of hypergeometric functions 
\begin{eqnarray}
R_s &=& (1+z)^{iQ} \big[{C_1} z^{-iQ} \nn \\
&&_{2}F_1(-l+s,l+1+s;1-\bar{Q}+s;-z) + {C_2} z^{iQ-s} \nn \\
&&_{2}F_1(-l+\bar{Q},l+1+\bar{Q};1+\bar{Q}-s;-z)\big] \,, \label{sol_nearhor}
\end{eqnarray}
where $\bar{Q}=2iQ$ and the integration constants ${C_1}$ and ${C_2}$ are related to the
amplitudes of outgoing and ingoing waves near the surface of the ECO, respectively. For $s=-2$, we transform the solution~\eqref{sol_nearhor} in the form given by Eq.~\eqref{DetweilerX}. The near-horizon behavior of the solution is given by Eq.~\eqref{Psip}, where the coefficients $B_{\text{out}}$ and $B_{\text{in}}$ are related to the integration constants $C_1$ and $C_2$, respectively.

The large-$r$ behavior of the solution~\eqref{sol_nearhor} is
\begin{eqnarray}
R_s &\sim& \left(\frac{r}{r_+ - r_-}\right)^{l-s} \Gamma(2l+1) \Bigg[\frac{{C_1} \ 
\Gamma(1-\bar{Q}+s)}{\Gamma(l+1-\bar{Q}) \Gamma(l+1+s)} \nn \\
&+& \frac{{C_2} \ \Gamma(1+\bar{Q}-s)}{\Gamma(l+1+\bar{Q}) \Gamma(l+1-s)} \Bigg] + 
\left(\frac{r}{r_+ - r_-}\right)^{-l-1-s} \nn \\
& & \frac{(-1)^{l+1+s}}{2 \Gamma(2l+2)} \Bigg[\frac{{C_1} \ \Gamma(l+1-s) 
\Gamma(1-\bar{Q}+s)}{\Gamma(-l-\bar{Q})} \nn \\
&+& \frac{{C_2} \ \Gamma(l+1+s) \Gamma(1+\bar{Q}-s)}{\Gamma(-l+\bar{Q})} \Bigg] \,, 
\label{Rhorizon}
\end{eqnarray}

At infinity, the radial wave equation~\eqref{wave_eq} for $M \omega \ll 1$ reduces to~\cite{Cardoso:2008kj}
\begin{equation}
r \partial_{r}^2 f_{s} + 2 (l+1-i \omega r) \partial_{r} f_{s} - 2 i (l+1-s) \omega f_{s} 
= 0 \,, \label{wave_eq_inf}
\end{equation}
where $f_{s} = e^{i \omega r} r^{-l+s} R_{s}$. The general solution of 
Eq.~\eqref{wave_eq_inf} is a linear combination of a confluent hypergeometric function 
and a Laguerre polynomial
\begin{eqnarray}
R_{s} &=& e^{-i \omega r} r^{l-s} \big[ C_3 \ U(l+1-s,2l+2,2i \omega r) \nn \\
&+& C_4 \ L_{-l-1+s}^{2l+1}(2 i \omega r) \big] \,,
\label{sol_infinity}
\end{eqnarray}
where the absence of ingoing waves at infinity implies $C_4 = (-1)^{l-s} \ C_3 \ \Gamma(-l+s)$. For $s=-2$, the solution~\eqref{sol_infinity} is turned in the form given by Eq.~\eqref{DetweilerX}.
In order to have a purely outgoing wave with unitary amplitude at infinity, as in Eq.~\eqref{Psip}, we impose
\begin{equation}
C_3 = \frac{(-i \omega)^{1+l} \ 2^l \ \Gamma(3+l)}{{\lambda}_{-2} \ {\lambda}_0 \ \Gamma(-1+l)} \,.
\end{equation}
The small-$r$ behavior of the solution~\eqref{sol_infinity} is
\begin{eqnarray}
R_{s} &\sim& C_3 \ r^{l-s} \frac{(-1)^{l-s}}{2} \frac{\Gamma(l+1+s)}{\Gamma(2l+2)} 
\nn \\
&+& C_3 \ r^{-l-1-s} (2 i \omega)^{-(2l+1)} \frac{\Gamma(2l+1)}{\Gamma(l+1-s)} \,. 
\label{Rinfinity}
\end{eqnarray}

The matching of Eqs.~\eqref{Rhorizon} and \eqref{Rinfinity} in the intermediate region 
yields
\begin{equation}
\frac{{C_1}}{{C_2}} = -\frac{\Gamma(l+1+s)}{\Gamma(l+1-s)} \left[\frac{R_+ + i (-1)^l 
(\omega (r_+ - r_-))^{2l+1} L S_+}{R_- + i (-1)^l (\omega (r_+ - r_-))^{2l+1} L S_-} 
\right] \label{eq:ab1}
\end{equation}
where
\begin{eqnarray}
R_\pm &\equiv& \frac{\Gamma(1 \pm \bar{Q} \mp s)}{\Gamma(l + 1 \pm \bar{Q})}\,, \quad
S_\pm \equiv \frac{\Gamma(1 \pm \bar{Q} \mp s)}{\Gamma(-l \pm \bar{Q})}\,, \nn \\
L &\equiv& \frac{1}{2} \left[\frac{2^l \, \Gamma(l+1+s)\Gamma(l+1-s)}{\Gamma(2l+1) 
\Gamma(2l+2)} \right]^{2} \,.
\end{eqnarray}
The reflection coefficient ${\cal R}_{\rm BH}=B_{\text{in}}/B_{\text{out}}$ is computed in terms of $C_2/C_1$. By using Eq.~\eqref{eq:ab1}, we derive an analytical expression for 
${\cal R}_{\rm BH}$ at low frequencies. For $l=2$, the equation for ${\cal R}_{\rm BH}$
reads 
\begin{widetext}
\begin{eqnarray}
\nn {\cal R}_{\rm BH}^{\rm LF} = -8 M k e^{\frac{\zeta (\gamma-1)}{\gamma+1}} \frac{2 M k 
-i(\gamma-1)}{(\gamma-1)^2} \left[ \frac{-M (\gamma-1) \xi}{L}\right]^{\zeta (\gamma-1)} \left[ \frac{16 k^2 
M^2}{(\gamma-1)^2}+1 \right] \times \\
 \frac{\Gamma(-2+\zeta) \Gamma(-1-\zeta) \left[ 1800 i \Gamma(-2-\zeta) + \left(\omega M (\gamma-1) \xi\right)^5 
\Gamma(3-\zeta) \right]}{\Gamma(-2-\zeta) \Gamma(3-\zeta) \left[ 1800 i \Gamma(-2+\zeta) + \left(\omega M (\gamma-1) 
\xi\right)^5 \Gamma(3+\zeta) \right]} \,,
\end{eqnarray}
\end{widetext}
where $\gamma=r_-/r_+$, $\xi = 1+\sqrt{1-\chi^2}$, $\zeta=i(2 \omega M - m \sqrt{\gamma})(\gamma+1)\xi/(\gamma-1)$, and 
$L$ is an arbitrary constant (with dimensions of a length) which is related to the integration constant of 
Eq.~\eqref{tortoise}. The expression of ${\cal R}_{\rm BH}$ is provided in a publicly available {\scshape 
Mathematica}\textsuperscript{\textregistered} notebook~\cite{webpage}.

\section{BH response at the horizon in some particular cases} \label{app:response}
In this appendix we provide some particular case for the BH response at the horizon, $Z_{\rm BH}^-$, for some
specific toy models of the source. We assume the latter is localized within the cavity. 

%

The simplest case is that of a source localized in space, and for which the frequency dependence can be factored out:
\begin{equation}
 \tilde S(\omega,x) = C(\omega) \exp\left(-(x-x_s)^2/\sigma^2\right)\,,
\end{equation}
where $|x_s|\ll M$. In this case, it is easy to show that 
\begin{equation}
 \tilde {\cal Z}_{\rm BH}^+= e^{2ik x_s} \tilde Z_{\rm BH}^+ \,. \label{eq:calZlocsource}
\end{equation}
This, together with Eq.~\eqref{ZBHminusTOT}, yields
\begin{eqnarray}
 \tilde Z_{\rm BH}^- = \left(\frac{e^{2ik x_s}+{\cal R}_{\rm BH}}{{\cal T}_{\rm BH}} \right)\tilde Z_{\rm BH}^+\,. 
\label{eq:bhtemplateHOR} 
\end{eqnarray}
Remarkably, the above relation is independent of the width of the Gaussian source $\sigma$ and of the function 
$C(\omega)$ characterizing the source, and it is also valid for any spin. Note that the above result is formally 
equivalent to the case of localized source studied in Ref.~\cite{Testa:2018bzd}, and in fact reduces to it when 
$\sigma\to0$ and $x_s$ coincides with the surface location $x_0$. 


Inspired by Eq.~\eqref{eq:calZlocsource}, one 
could also parametrize the BH response $\tilde {\cal Z}_{\rm BH}^+$ relative to $\tilde {Z}_{\rm BH}^+$ in a 
model-agnostic way with a generic (complex) proportionality factor:
\begin{equation}
 \tilde {\cal Z}_{\rm BH}^+ = \eta e^{i\nu} Z_{\rm BH}^{+}\,,
\end{equation}
where $\eta$ and $\nu$ are (real) parameters of the template.
Since the BH response is dominated by the QNMs, a model in which $\tilde {\cal Z}_{\rm BH}^+ = {\cal F}(\omega) Z_{\rm 
BH}^{+}$ can be effectively reduced to $\tilde {\cal Z}_{\rm BH}^+ = {\cal F}(\omega_R) Z_{\rm 
BH}^{+}$. In such case the term ${\cal F}(\omega_R)=\eta e^{i\nu}$ is a generic parametrization of a 
complex number. 


Finally, another possible model is to consider a plane-wave source that travels towards $\pm\infty$, in this case we 
have
\begin{eqnarray}
\tilde S(x, \omega) &=&
 \int dt e^{i \omega t} S(x, t)\nonumber\\
 &=&\int dt e^{i \omega t} S(0, t\mp x) =\tilde S(0, \omega) e^{\pm i \omega x}\,.
\end{eqnarray}
Using Eq.~\eqref{ZBH}, we obtain
\begin{equation}\label{A7}
 \tilde Z_{\rm BH}^{+}(\omega) = \tilde Z_{\rm BH}^{-}(\omega) \frac{\int_{-\infty}^{+\infty}dx\Psi_{-} e^{\pm i \omega 
x}}{\int_{-\infty}^{+\infty}dx \Psi_{+} e^{\pm i \omega x}}\,,
\end{equation}
or, more explicitly
\begin{widetext}
\begin{eqnarray}
&&\tilde Z_{\rm BH}^{+}(\omega) = \tilde Z_{\rm BH}^{-}(\omega) \frac{\int_{{x \sim 0}}dx\Psi_{-} e^{i \omega x} 
+\int^{\infty}(A_{\textrm{out}}e^{2i \omega 
		x} +A_{\textrm{in}})dx +\int_{-\infty}dx e^{i m \Omega x} }{\int_{x \sim 0}dx\Psi_{+} e^{i \omega x}
	+\int^{\infty} e^{2i \omega x} dx +\int_{-\infty}(B_{\textrm{out}} e^{2i \omega x-i m \Omega 
x}+B_{\textrm{in}} e^{i m \Omega x} )dx \,,
}\nonumber\,,
 \end{eqnarray}
\end{widetext}
where ${x \sim 0}$ is the region where the potential is non-zero and we considered only the upper-sign case for 
ease of notation. Considering that $\tilde Z_{\rm BH}^{+}(\omega)$ has a pole at 
$\omega_{\rm QNM}=\omega_R+i\omega_I$ we expect also $\tilde Z_{\rm BH}^{-}(\omega)$ to have such a pole.
Since $\Im\omega_{\rm QNM}<0$ the terms $\int^{+\infty}dx$ dominate the numerator and the denominator for $\omega 
\approx \omega_{\rm QNM}$, and we obtain
\begin{equation}\label{A9}
\tilde Z_{\rm BH}^{+} \approx -\left(\frac{\mathcal{R}_{\rm BH}}{\mathcal{T}_{\rm BH}}\right)^* \tilde Z_{\rm BH}^{-}\,.
\end{equation}
The case with the lower sign (plane wave traveling toward $-\infty$) gives the same result.
%
\bibliographystyle{utphys}
\bibliography{Ref}
\end{document}